\documentclass[aps,pre,reprint,superscriptaddress,nofootinbib]{revtex4-2}

\usepackage{amsmath,amssymb,bm}
\usepackage{graphicx}
\usepackage{hyperref}
\usepackage{cleveref}
\usepackage{algorithm}
\usepackage[noend]{algpseudocode}
\usepackage{booktabs}
\usepackage{xcolor}
\usepackage{makecell}
\usepackage[colorinlistoftodos]{todonotes}
\usepackage[english]{babel}
\usepackage{underscore}
\usepackage{pbalance}

\hypersetup{colorlinks=true,linkcolor=blue,citecolor=blue,urlcolor=blue}

\newcommand{\paren}[1]{\left( #1 \right)}

\newcommand{\brackets}[1]{\left\langle #1 \right\rangle}
\newcommand{\squarebrackets}[1]{\left[ #1 \right]}

\definecolor{lavender}{rgb}{0.858, 0.688, 0.878}
\definecolor{lightblue}{rgb}{0.678, 0.847, 0.902}
\definecolor{lightgreen}{rgb}{0.678, 0.847, 0.678}

\newcommand{\at}[2]{#1_{#2}}
\newcommand{\atalg}[2]{#1^{[#2]}}
\newcommand{\hypergraph}{\mathcal{H}}
\newcommand{\nodes}{\mathcal{N}}
\newcommand{\edges}{\mathcal{E}}
\newcommand{\newedge}{e}
\newcommand{\seededge}{f}
\newcommand{\focalnode}{u}
\newcommand{\node}{v}

\newcommand{\extantnodesinedge}{X}
\newcommand{\novelnodesinedge}{Y}
\newcommand{\jointdist}{q}

\newcommand{\vz}{\mathbf{z}}

\newcommand{\paramvector}{\boldsymbol{\theta}}
\newcommand{\param}{\theta}

\newcommand{\expectedcopiednodes}{\nu}

\newcommand{\copyrate}{\rho}
\newcommand{\extantrate}{\gamma}
\newcommand{\novelrate}{\eta}
\newcommand{\same}[1]{#1_+}
\newcommand{\opp}[1]{#1_-}

\newcommand{\abs}[1]{\left| #1 \right|}

\newcommand{\sufficientstatsvec}{\mathbf{s}}
\newcommand{\lr}{\alpha} 
\newcommand{\vzero}{\mathbf{0}}

\newcommand{\argmax}{\operatorname*{argmax}}

\newcommand{\epoch}{\ell}
\newcommand{\step}{\tau}

\newcommand{\Model}{Copying Hyperedges Influenced by Label Interactions}
\newcommand{\model}{CHILI}

\newcommand{\cL}{\mathcal{L}}

\newcommand{\poisson}[1]{\mathrm{Poisson}\paren{#1}}
\newcommand{\poissonpdf}[2]{\frac{#1^{#2} e^{-#1}}{#2!}}

\newcommand{\lik}{p}
\newcommand{\acceptprob}{a}
\newcommand{\vx}{\mathbf{x}}
\newcommand{\vy}{\mathbf{y}}

\newcommand{\moment}{\mu}

\newcommand{\maxepochs}{E}

\newcommand{\R}{\mathbb{R}}
\newcommand{\vdelta}{\boldsymbol{\delta}}

\newcommand{\vpsi}{\boldsymbol{\psi}}

\newcommand{\simannviznumsteps}{1340}
\newcommand{\simannviznumnodes}{67}
\newcommand{\simannviznumedges}{200}
\newcommand{\simannvizmaxll}{-10.38}
\newcommand{\simannvizmaxllari}{0.39}

\newcommand{\simannvizspectralari}{-0.03}
\newcommand{\simannvizsamecopyrate}{0.90}
\newcommand{\simannvizoppcopyrate}{0.10}
\newcommand{\simannvizsameextantrate}{1.00}
\newcommand{\simannvizoppextantrate}{0.25}
\newcommand{\simannvizsamenovelrate}{0.20}
\newcommand{\simannvizoppnovelrate}{0.10}

\newcommand{\thetaguess}{(0.9, 0.1, 0.2, 0.1, 2.0, 0.3)}

\newcommand{\synthfixedetaplus}{0.8}
\newcommand{\synthfixedetaminus}{0.15}
\newcommand{\synthfixedgammaplus}{0.7}
\newcommand{\synthfixedgammaminus}{0.2}
\newcommand{\synthfixedlambdaplus}{0.8}
\newcommand{\synthfixedlambdaminus}{0.5}

\begin{document}

\title{Growing Hypergraphs with Homophily}
\author{Violet Ross}
\thanks{Equal author contributions.}

\affiliation{Department of Computer Science, University of Colorado at Boulder, Boulder, CO, USA}
\author{Francis Cataldo}
\thanks{Equal author contributions.}

\affiliation{Department of Computer Science, Middlebury College, Middlebury, VT, USA}
\author{Philip S. Chodrow}
\affiliation{Department of Computer Science, Middlebury College, Middlebury, VT, USA}

\date{\today}
\begin{abstract}
    There are many extant models of hypergraphs with interactions governed by attribute-based homophily between nodes, but most of these models assume independence between edges conditional on node parameters. 
    Relaxing this assumption, we study a mechanistic model of growing hypergraphs in which edge formation is influenced by both previous edges and binary node labels. 
    Edges form in this model as noisy copies of previous edges, where the transmission of nodes from one edge to the next depends on multiple homophilic mechanisms governed by labels.
    These mechanisms give rise to tunable assortative structure in the hypergraph. 
    We first derive several asymptotic properties of our model, including the joint distribution of labels within edges and the degree distribution.
    We then turn to two inference problems. 
    Our model defines a likelihood over a labeled hypergraph, allowing us to use standard maximum-likelihood techniques to structure algorithms. 
    We estimate the model parameters on synthetic and real data via (stochastic) expectation maximization. 
    These estimates give statistically-principled descriptions of the operation of homophily in empirical polyadic systems. 
    We also demonstrate an approach to community detection via simulated annealing which, though computationally expensive, achieves competitive results on both synthetic data and certain empirical data sets known to be challenging to community detection techniques based on the edge-independence assumption. 
    Our findings highlight the importance of incorporating edge- and label-dependence in higher-order modeling and data analysis. 
\end{abstract}

\maketitle

\section{Introduction}

    Hypergraphs provide a natural, flexible data representation for complex systems with polyadic (multi-way) interactions; such systems form one class of \emph{higher-order networks} \cite{battistonNetworksPairwiseInteractions2020,bick2023higher}.
    Systems naturally modeled as hypergraphs include social interaction networks (both in-person \cite{mastrandrea2015contact} and online \cite{bensonSimplicialClosureHigherorder2018}), academic co-authorship neteworks, \cite{wu2022hypergraph}, and networks of ecological relations \cite{levine2017beyond}. 
    Much recent work has emphasized the importance of faithfully representing polyadic structure rather than reducing it to dyadic interactions via graph projections \cite{battistonNetworksPairwiseInteractions2020,bick2023higher} .

    Our focus in this work is the mechanism of homophily---the tendency of nodes to interact with other nodes with similar attributes---in hypergraph data. 
    One way to model homophily in hypergraphs is via stochastic generative models, which define probability distributions over hypergraphs. 
    In one standard framework, a set $\nodes$ of $n$ nodes is initialized along with a vector $\vz \in \{1,\ldots,k\}^n$ that assigns to each node one of $k$ community labels. 
    One then forms the edge set $\edges$ iteratively by visiting each subset $R \subseteq \nodes$ and independently assigning $a_R \in \mathbb{N}$ (possibly repeated) edges to $R$ according to a distribution $\lik(a_R | \vz_R ;\paramvector)$, where $\vz_R$ is the vector of labels of the nodes in $R$ and $\paramvector$ is a set of model parameters.
    Independence of sampling results in a probability distribution over hypergraphs which can be factorized into a product of edge probabilities conditioned on the node labels and model parameters: 
    \begin{align}
        \lik(\hypergraph | \vz; \paramvector) = \prod_{R \subseteq \nodes} \lik(a_R | \vz_R; \paramvector) \;. \label{eq:edge-independence}
    \end{align}
    Many generative models fall into this broad scheme, including hypergraph stochastic block models and some of their generalizations \cite{chodrowGenerativeHypergraphClustering2021,ruggeriCommunityDetectionLarge2023,kim2018stochastic,badalyanStructureInferenceHypergraphs2024,hood2026broad}.

    Models of this type often underly algorithms for hypergraph community detection (also called \emph{clustering} or \emph{partitioning}), in which one seeks to estimate the latent node labels $\vz$ given an observed hypergraph $\hypergraph$.
    The conditional independence property \eqref{eq:edge-independence} is extremely useful for computation in this context.  
    Factorizability of the likelihood enables many efficiencies in various inference algorithms, including approximate maximum-likelihood estimation via direct optimization \cite{chodrowGenerativeHypergraphClustering2021} or expectation-maximization \cite{ruggeriCommunityDetectionLarge2023}; spectral methods \cite{chodrowNonbacktrackingSpectralClustering2023,fernandez2026achieving,li2026higher}; and message passing \cite{ruggeri2024message}. 
    Another recent optimization-based approach \cite{kritschgauCommunityDetectionHypergraphs2024} is equivalent to a microcanonical version of the stochastic block model, which satisfies an asymptotic form of edge-independence in the large sparse limit.
    Latent-space models (e.g. \cite{yuModelingHypergraphsDiversity2025}) also satisfy a form of edge independence, albeit in the context of a very different stochastic generating process.
    
    Edge independence, however, is a strong and often unrealistic assumption for modeling hypergraph data. 
    A parallel line of work has found that polyadic data sets often contain significant edge correlations \cite{bensonSimplicialClosureHigherorder2018,landrySimplicialityHigherorderNetworks2024,leeHowHyperedgesOverlap2021}.
    These correlations are often observed as \emph{similarity between interactions}.
    When an interaction occurs on a set $R \subseteq \nodes$, this increases the likelihood of observing a new interaction on a set $S \subseteq \nodes$ correlated with $R$ in the sense that $R\cap S$ is larger than would be expected by chance. 
    That is, new interactions are likely to have large overlap with the interactions we have already observed.
    These overlaps can be shown to affect the kinds of dynamics which can unfold on the hypergraph \cite{maliziaHyperedgeOverlapDrives2025,lamata2025hyperedge}. 
    Several generative models for the growth of hypergraphs with edge-dependence have been proposed \cite{bensonSequencesSets2018,avinRandomPreferentialAttachment2019,heEdgeCorrelationsLink2025,leeTHyMeTemporalHypergraph2021,leeTemporalHypergraphMotifs2023}, including some with explicit nodel labels and corresponding homophilic mechanisms \cite{giroirePreferentialAttachmentHypergraph2022,giroirePreferentialAttachmentHypergraph2022a,kaminski2023hypergraph}; see \cite{laber2026guide} for a recent review.\footnote{We note that the above are all models \emph{of} hypergraph formation rather than models of dynamics \emph{on} hypergraphs; recent work argues that the latter may be better expressed via dyadic graph models \cite{peixoto2026graphs} but does not treat the former.}

    Of the generative models of hypergraph growth with both edge-dependence and node labels, we are unaware of any which allow for statistically-motivated inference of either the dynamical parameters or the latent node labels.
    In this paper, we aim to connect stochastic generative modeling of edge-dependent, node-labeled hypergraphs to inference algorithms. 
    To this end, we introduce a model of hypergraphs with binary node labels in which edges form via a noisy edge copy process tuned by homophilic interactions between the node labels.
    We refer to this model as \Model{} (\model{}).
    Our presentation of \model{} proceeds as follows.
    In \Cref{sec:model-description}, we describe \model{} and several of its limiting properties, including edge structure and degree distribution.
    We turn to optimization-based inference methods in \Cref{sec:inference}, considering both parameter inference and community detection.
    Evaluating the likelihood of our model requires marginalizing over a set of latent variables that govern the noisy copy mechanism, which is computationally expensive. 
    To estimate our model's parameters when node labels are known, we therefore describe a stochastic expectation-maximization algorithm with favorable scaling.
    Community detection in this model is much more computationally challenging. 
    As an early step in this direction,we describe a community detection algorithm for estimating node labels via simulated annealing on the \model{} likelihood. 
    In \Cref{sec:results}, we demonstrate our algorithms on synthetic and empirical data. 
    We show that, on synthetic data sampled from \model{}, our stochastic expectation maximization algorithm is both consistent and scalable, and that our simulated annealing algorithm for community detection, though computationally expensive, outperforms other standard methods in certain parameter regimes. 
    On empirical data, our parameter estimates yield insight into the mechanisms governing edge formation.
    \model{} community detection achieves competitive recovery of observed metadata labels on some data sets known to be challenging to other recent community detection methods.
    We conclude in \Cref{sec:discussion} with some discussion of future directions for models of node-labeled hypergraphs with edge-dependence. 
    Our findings highlight the benefits of using explicit stochastic generative models to guide data analysis algorithms for studying homophily and assortativity in polyadic systems.

\section{The \model{} Model} \label{sec:model-description}

    We propose \Model{} (\model{}), a generative model of growing hypergraphs with binary node labels. 
    \model{} is an extension of the Hyperedge Copy Model (HCM) of He et al. \cite{heEdgeCorrelationsLink2025}. 
    In the HCM, a new edge $\newedge$ is formed through a combination of three distinct steps: a copy step, in which a previous edge $\seededge$ is sampled and imperfectly copied into $\newedge$; an extant node addition step, in which nodes in the remainder of the node set, $\nodes \setminus \seededge$, are added to $\newedge$; and a novel node addition step, in which nodes are added to both $e$ and $\nodes$ for the first time.
    \model{} replicates these mechanisms, but adds label-aware parameters to regulate each. 

    Formally, a \model{} model state at time $t$ is a hypergraph $\at{\hypergraph}{t} = \paren{\at{\nodes}{t}, \at{\edges}{t}}$, where $\at{\nodes}{t}$ is the set of nodes and $\at{\edges}{t}$ is the set of hyperedges, as well as a binary label vector $\at{\vz}{t} \in \{0, 1\}^{|\at{\nodes}{t}|}$ which assigns to each node a label of either 0 or 1; $z_u$ denotes the label of node $u$.
    Let $\bar{z} = 1 - z$ denote the opposite label of $z$.
    At timestep $t+1$, we add edge $\newedge$ to $\at{\edges}{t}$ to obtain $\at{\edges}{t+1}$, where $\newedge$ is formed according to the following three-step process:
    \begin{itemize}
        \item 
            \textbf{Edge-Copying}: An edge $\seededge$, called the \emph{seed edge}, is sampled uniformly at random from $\at{\edges}{t}$, and a node $\focalnode$, called the \emph{focal node}, is sampled uniformly at random from $\seededge$. 
            The node $\focalnode$ is included in $\newedge$.
            Then, each node $\node$ in $\seededge \setminus \{\focalnode\}$ is included in $\newedge$ with probability $\same{\copyrate{}}$ if $z_{\node} = z_{\focalnode}$ and with probability $\opp{\copyrate{}}$ if $z_{\node} \neq z_{\focalnode}$.
        \item 
            \textbf{Extant Node Addition}: We sample $\extantnodesinedge_{z_\focalnode} \sim \poisson{\same{\extantrate{}}}$ nodes from $\at{\nodes}{t} \setminus \seededge$ with label $z_\focalnode$ and add them to $\newedge$. 
            We similarly sample $\extantnodesinedge_{\bar{z}_\focalnode} \sim \poisson{\opp{\extantrate{}}}$ nodes from $\at{\nodes}{t} \setminus \seededge$ with label $\bar{z}_\focalnode$ and add them to $\newedge$.
            In the event that there are not enough nodes with the required label in $\at{\nodes}{t} \setminus \seededge$, we add all available nodes with that label to $\newedge$.
        \item \textbf{Novel Node Addition}: We sample $\novelnodesinedge_{z_\focalnode} \sim \poisson{\same{\novelrate{}}}$ and add $\novelnodesinedge_{z_\focalnode}$ nodes with label $z_\focalnode$ to both $\newedge$ and $\at{\nodes}{t+1}$. 
        We similarly sample $\novelnodesinedge_{\bar{z}_\focalnode} \sim \poisson{\opp{\novelrate{}}}$ and add $\novelnodesinedge_{\bar{z}_\focalnode}$ nodes with label $\bar{z}_\focalnode$ to both $\newedge$ and $\at{\nodes}{t+1}$. 
        The associated labels are also added to $\at{\vz}{t}$ to form $\at{\vz}{t+1}$.
    \end{itemize}
    We let $\paramvector = (\same{\copyrate{}}, \opp{\copyrate{}}, \same{\extantrate{}}, \opp{\extantrate{}}, \same{\novelrate{}}, \opp{\novelrate{}})$ denote the complete vector of model parameters. 
    Formally, the probability of realizing a given hypergraph $\hypergraph$ with label vector $\vz$ under \model{} at time $t$ can be described as a time-dependent probability distribution $\at{p}{t}(\hypergraph, \vz ; \paramvector)$ over the space of hypergraphs with labeled nodes; we omit the time index when otherwise clear from context. 
    Our \model{} model leverages the hypergraph data structure supplied by the XGI package for polyadic data analysis in Python \cite{landryXGIPythonPackage2023}. 

\begin{figure}
    \includegraphics[width=0.495\textwidth]{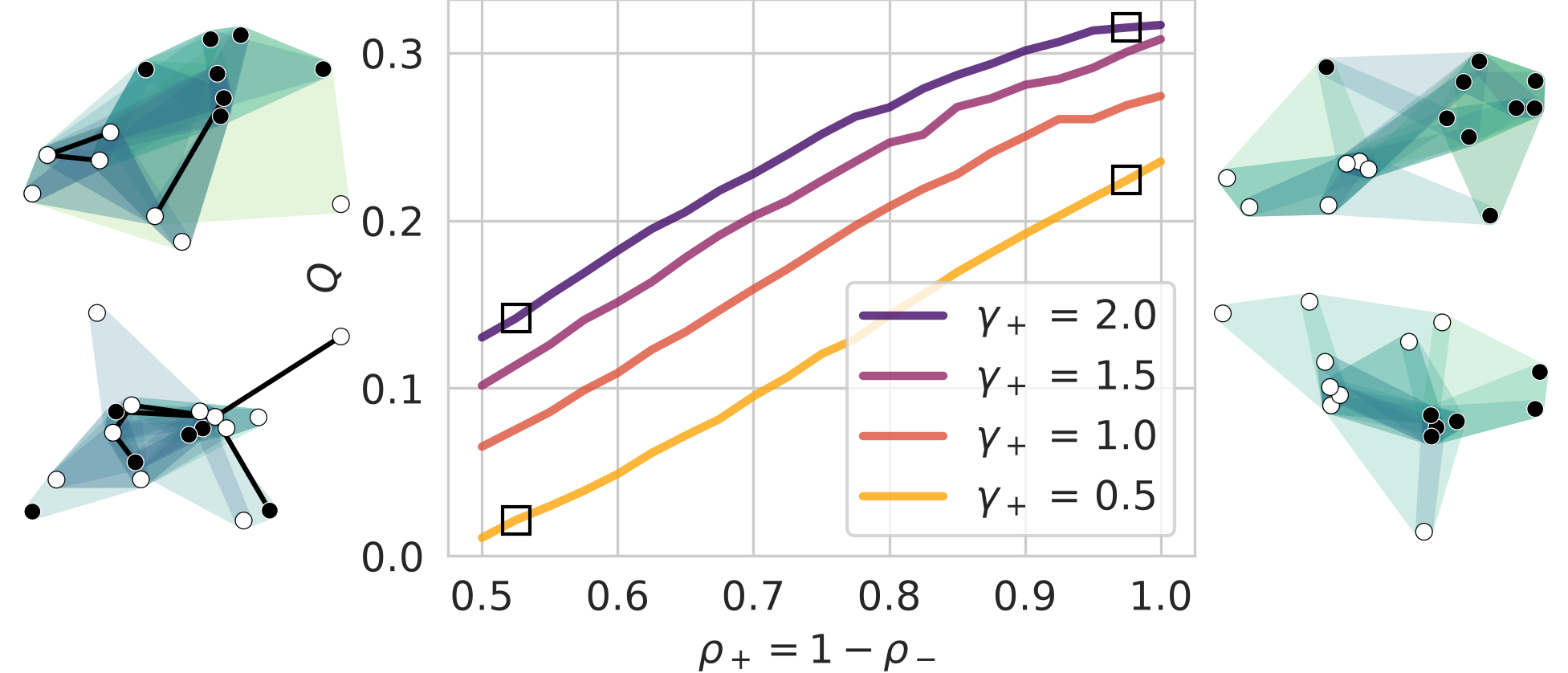}
        \caption{
            Suite of model simulations illustrating the dependence of community structure, as measured by clique-projection modularity $Q$, on the \model{} parameters. 
            The parameters $\opp{\extantrate} = 0.5$ $\same{\novelrate{}} = 0.2$, and $\opp{\novelrate{}} = 0.1$ are held fixed. 
            Each line corresponds to a different value of $\same{\extantrate{}}$ indicated by the legend.
            Along the horizontal axis, we vary $\same{\copyrate{}}$ and set $\opp{\copyrate{}} = 1 - \same{\copyrate{}}$. 
            Each parameter combination is simulated $100$ times for $1,000$ timesteps. 
            On either side of the plot, we show small hypergraphs sampled from parameter combinations indicated by the square markers for illustration.
        }\label{fig:community-structure-illustration}
    \end{figure}

    \Cref{fig:community-structure-illustration} illustrates the role of the model parameters in tuning the community structure of realized hypergraphs. 
    We measure community structure by computing Newman-Girvan modularity \cite{newman2004finding} on the weighted clique-projection of the hypergraph. 
    We also show several sample hypergraphs at different parameter combinations for visual illustration. 
    For fixed $\opp{\extantrate}$, $\same{\novelrate{}}$, and $\opp{\novelrate{}}$, increasing the difference $\same{\copyrate{}} - \opp{\copyrate{}}$ leads to higher modularity, as does increasing $\same{\extantrate{}}$.
    Hypergraphs sampled from our model using high values of $\same{\copyrate{}}$ and $\same{\extantrate{}}$ tend to have label-homogeneous cores of densely overlapping hyperedges, which are weakly connected to one another by edges with more heterogeneous label compositions.
    Hypergraphs with lower values of $\same{\copyrate{}}$ and $\same{\extantrate{}}$ tend to have less-structured distributions of labels over hyperedges. 
    We note that the \model{} model can also be used to generate hypergraphs under heterophilic conditions, though we do not explore this regime further here. 
    We also note that the \model{} model does not incorporate any explicit dependence of homophily on edge size, though such a dependence could be incorporated in future work.

\subsection{Asymptotic Properties} \label{sec:asymptotics}
    
    In order to contextualize the data analysis algorithms that we develop in \Cref{sec:inference}, we describe some of the macroscopic behaviors of the \model{} model in the long-time limit. 
    As we show, the \model{} model asymptotically approaches a stationary regime in which (a) equal numbers of nodes of each label are present in the hypergraph and (b) the degree distribution is approximately power-law with an exponent that can be computed in closed form from the model parameters.
    These considerations suggest that our inferential methods may be most appropriate for data sets which share these features.

    \subsubsection{Joint Distribution of Labels in Edges} \label{sec:joint-distribution}

    \begin{figure*}[t]
        \centering
        \includegraphics[width=1\textwidth]{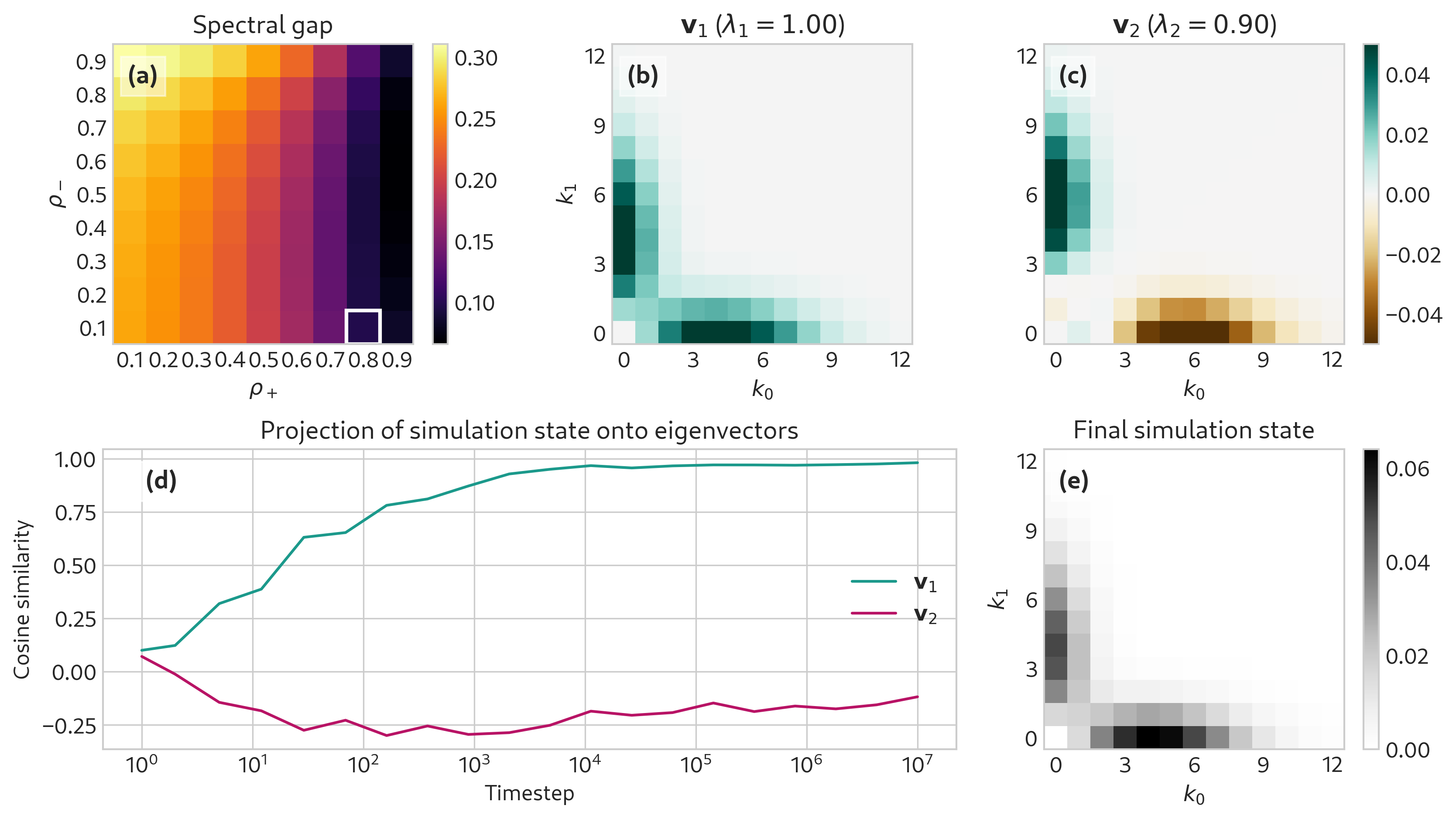}
        \caption{
            Illustrative long-run dynamics of our model.
            (a): Spectral gap $1 - \lambda_2$ of the linear map $T$ as a function of $\same{\copyrate{}}$ and $\opp{\copyrate{}}$ for fixed parameter values $\same{\extantrate{}} = 0.5$, $\opp{\extantrate{}} = 0.2$, $\same{\novelrate{}} = 0.4$ and $\opp{\novelrate{}} = 0.2$. 
            The highlighted values $\same{\copyrate{}} = 0.8$ and $\opp{\copyrate{}} = 0.1$ are used in the remainder of the visualization. 
            (b): Leading eigenvector $\mathbf{v}_1$ of $T$, describing the stationary joint distribution $\at{\jointdist}{\infty} \paren{k_0, k_1}$ of label counts in edges.
            (c): Second eigenvector $\mathbf{v}_2$ of $T$. 
            The transients of the model dynamics are dominated by the subspace spanned by $\mathbf{v}_1$ and $\mathbf{v}_2$. 
            (d): Euclidean projections of the simulated joint distribution $\at{\jointdist}{t} \paren{k_0, k_1}$ onto the eigenvectors $\mathbf{v}_1$ and $\mathbf{v}_2$ during a simulation run with $10^7$ timesteps. 
            At each time point, we aggregate the most recent $1,000$ edges to compute the joint distribution in order to approximate an ``instantaneous'' simulation state. 
            (e): Empirical joint distribution $\at{\jointdist}{t} \paren{k_0, k_1}$ aggregated across $10^7$ timesteps. 
        }
        \label{fig:dynamics-eigenvectors}
    \end{figure*}

    Several asymptotic properties of our \model{} model can be described in terms of the joint distribution $\jointdist(k_0, k_1)$ of node labels in an edge; here $\jointdist(k_0, k_1)$ gives the probability that an edge selected uniformly at random from $\edges$ contains exactly $k_0$ nodes of label 0 and $k_1$ nodes of label 1.
    Under the $t\rightarrow \infty$ limit this joint distribution can be obtained as the leading eigenvector of a linear map. 
    Let $\at{\jointdist}{t}(k_0, k_1)$ be the joint distribution of label counts in a uniformly random edge after $t$ timesteps, and let $\jointdist(k_0, k_1) = \at{\jointdist}{\infty} \paren{k_0, k_1}$ be the stationary distribution.
    Then, the probability $r(k_0', k_1')$ of realizing an edge with $k_0'$ nodes of label 0 and $k_1'$ nodes of label 1 in timestep $t+1$ is given by 
    \begin{align*}
        r(k_0', k_1') = \sum_{k_0, k_1} T(k_0', k_1' | k_0, k_1)\at{\jointdist}{t}(k_0, k_1)  \;,
    \end{align*}
    where $T(k_0', k_1' | k_0, k_1)$ gives the probability that the edge $e$ realized in step $t+1$ has $k_0'$ nodes of label 0 and $k_1'$ nodes of label 1 given that the edge $f$ sampled to form $e$ has $k_0$ nodes of label 0 and $k_1$ nodes of label 1.
    At stationarity, we must have $r(k_0', k_1') = \at{\jointdist}{t}(k_0', k_1') = \jointdist \paren{k_0', k_1'}$, so the stationarity condition can be obtained by solving the linear system 
    \begin{align*}
        \jointdist \paren{k_0', k_1'} = \sum_{k_0, k_1} T(k_0', k_1' | k_0, k_1)\jointdist \paren{k_0, k_1}  \;.
    \end{align*}
    The stationary joint distribution $\jointdist \paren{k_0, k_1}$ is therefore given by the leading eigenvector of the doubly-indexed matrix $T$ with entries $T(k_0', k_1' | k_0, k_1)$. 
    The entries of this matrix follow from the parametric probability descriptions used to define the \model{} model.
    Conditional on choosing an edge $\seededge$ with $k_0$ nodes of label 0 and $k_1$ nodes of label 1, a node of label $z$ is selected as the focal node $\focalnode$ with probability $\frac{k_z}{k_0 + k_1}$.
    Then, the number of nodes of labels $z$ and $\bar{z}$ in $\newedge$ have distributions
    \begin{align*}
        k_z' &\sim 1 + \text{Binom}\paren{k_z - 1, \same{\copyrate{}}} + \text{Poisson}(\same{\extantrate{}} + \same{\novelrate{}}) \;, \\
        k_{\bar{z}}' &\sim \text{Binom}\paren{k_{\bar{z}}, \opp{\copyrate{}}} + \text{Poisson}(\opp{\extantrate{}} + \opp{\novelrate{}}) \;.
    \end{align*}
    The probability of obtaining $k_0'$ nodes of label 0, conditioning on $k_0$ and $k_1$, is therefore 
    \begin{widetext}
        \begin{align*}
            t(k_0' | k_0, k_1) =& \frac{k_0}{k_0 + k_1} \sum_{i=0}^{k_0' - 1} \binom{k_0 - 1}{i} \same{\copyrate{}}^i (1 - \same{\copyrate{}})^{k_0 - 1 - i} 
            e^{-(\same{\extantrate{}} + \same{\novelrate{}})} \frac{(\same{\extantrate{}} + \same{\novelrate{}})^{k_0' - 1 - i}}{(k_0' - 1 - i)!}  + \\ 
            & \frac{k_1}{k_0 + k_1} \sum_{i=0}^{k_0'} \binom{k_0}{i} \opp{\copyrate{}}^i (1 - \opp{\copyrate{}})^{k_0 - i} 
            e^{-(\opp{\extantrate{}} + \opp{\novelrate{}})} \frac{(\opp{\extantrate{}} + \opp{\novelrate{}})^{k_0' - i}}{(k_0' - i)!}\;.
        \end{align*}
    \end{widetext}
    A parallel expression describes $t(k_1' | k_0, k_1)$, and the transition matrix entries are given by the product 
    \begin{align*}
        T(k_0', k_1' | k_0, k_1) = t(k_0' | k_0, k_1) t(k_1' | k_0, k_1) \;.
    \end{align*}
    In \Cref{fig:dynamics-eigenvectors}, we study the dynamics of our \model{} model as described by the matrix $T$. 
    The spectral gap $1 - \lambda_2$ of $T$ is a measure of the rate of convergence to the stationary distribution $\jointdist \paren{k_0, k_1}$, where $\lambda_2$ is the second-largest eigenvalue of $T$.
    For fixed values of the parameters $\same{\extantrate{}}$, $\opp{\extantrate{}}$, $\same{\novelrate{}}$ and $\opp{\novelrate{}}$, the spectral gap $1 - \lambda_2$ of $T$ is highest when $\opp{\copyrate}$ is large and $\same{\copyrate}$ is small. 
    High values of the spectral gap correspond to fast convergence to the stationary distribution $\jointdist \paren{k_0, k_1}$ given by the Perron eigenvector $\mathbf{v}_1$ of $T$ with eigenvalue $1$ (panel (b)). 
    We instead illustrate an alternative regime of relatively slow convergence in which $\same{\copyrate}$ is large and $\opp{\copyrate}$ is small.
    In this regime, the simulation may maintain transient dynamics in the direction of the second eigenvector $\mathbf{v}_2$ of $T$ with eigenvalue $\lambda_2 < 1$ (panel (c)) for relatively long periods of time before converging to the stationary distribution. 
    We illustrate this behavior in panel (d), where the system converges relatively slowly to the leading eigenvector $\mathbf{v}_1$ and away from the transient component $\mathbf{v}_2$. 
    The resulting final simulation state matches the stationary distribution $\jointdist \paren{k_0, k_1}$ relatively closely (panel (e)).

    \Cref{fig:dynamics-eigenvectors} shows that, when $\same{\copyrate{}}$ is sufficiently large, the stationary distribution $\jointdist \paren{k_0, k_1}$ is concentrated on hypergraphs in which typical edges have large majorities of nodes with the same label (panel b), the signature of higher-order structural assortativity.
    This distribution is symmetric with respect to the two node labels, however, indicating that although local edges may have dominant majority labels, the long-run dynamics converge to equal global label populations.  
    The second eigenvector $v_2$ (panel c), however,  is antisymmetric, describing a transient in which a single label group tends to dominate most edges. 
    When $\same{\copyrate{}}$ is sufficiently large, the spectral gap $1 - \lambda_2$ is small (panel a), indicating that this transient decays slowly and that the system may be dominated by edges with a single majority label for relatively long periods of time before equilibrating to the leading eigenvector (panels (d-e)).

    The presence of long transients suggests that, although the \model{} model ultimately converges to a stationary regime with power-law degrees and balanced label populations, it may also serve through its transients as a reasonable generative model of systems without these features. 
    We note that in settings it may be an undesirable feature of models to possess long transients; in our setting, since we are especially interested in statistical inference as defined below, the long transients in the dynamics do not pose a direct problem.  

\subsubsection{Degree Distribution} \label{sec:degree-distribution}

    One model feature we can obtain from the stationary joint distribution $\jointdist \paren{k_0, k_1}$ is the asymptotic degree distribution generated from \model{}.
    This asymptotic distribution follows a power law with an exponent which can be computed in closed form from the model parameters and the stationary joint distribution $\jointdist \paren{k_0, k_1}$. 
    Our derivation is similar to that of He et al. \cite{heEdgeCorrelationsLink2025}, which in turn is based on a derivation by Mitzenmacher \cite{mitzenmacherBriefHistoryGenerative2004}. 
    The argument requires a mean-field assumption. 
    Let $k = k_0 + k_1$ be the size of a uniformly random edge. 
    We assume that, when a node $\focalnode$ is selected from edge $\seededge$ in the edge-sampling step, the degree $d$ of $\focalnode$ is independent of the ratio $\frac{k_0}{k}$ of label-0 nodes in the edge $\seededge$ sampled to form the new edge $\newedge$.
    Under this assumption, it is possible to show that the degree distribution follows a power law with exponent
    \begin{align}
        \zeta &= 1 + \frac{\brackets{k_0} + \brackets{k_1}}{1 + \same{\copyrate{}}(\moment_{00} + \moment_{11} - 1) + 2\opp{\copyrate{}} \moment_{01}} \label{eq:power-law-exponent}\\ 
        \moment_{ij} &= \brackets{\frac{k_i k_j}{k}} \;. \nonumber
    \end{align}
    In the expression for $\moment_{ij}$ (and for $\brackets{k_i}$) the expectation is taken with respect to the stationary joint distribution $\jointdist \paren{k_0, k_1}$.
    A derivation of this result and a computational illustration are provided in Appendix \Cref{sec:degree-distribution-appendix} and \Cref{fig:degrees}. 
    
\section{Model Inference} \label{sec:inference}

    In this section, we consider the problems of \emph{parameter inference} (learning $\paramvector$ given $\hypergraph$ and $\vz$) and \emph{community detection} (learning $\vz$ given $\hypergraph$ and $\paramvector$).

\subsection{Parameter Estimation via Stochastic EM} \label{sec:sem}

    We first consider the problem of learning the parameter vector $\paramvector = (\same{\copyrate{}}, \opp{\copyrate{}}, \same{\extantrate{}}, \opp{\extantrate{}}, \same{\novelrate{}}, \opp{\novelrate{}})$ given an observed hypergraph $\hypergraph$ with sequential edge indices and known node labels $\vz$.
    We aim to do this via maximum-likelihood estimation. 
    Computing the complete data likelihood requires marginalizing over the model latent variables, which in this case are the edge $f$ and node $u$ selected at each timestep of the growth model to form the new edge $e$. 
    For $\newedge \in \edges$, let $t_\newedge$ denote the time at which $\newedge$ was added to the hypergraph.
    We say that $\seededge \prec \newedge$ if $t_{\seededge} < t_{\newedge}$. 
    Then, the complete data likelihood can be written 
    \begin{widetext}
    \begin{align}
        \lik(\hypergraph, \vz ; \paramvector) = \prod_{\newedge \in \edges} \sum_{\seededge \prec \newedge} \sum_{\focalnode \in \seededge} \lik(\newedge, \vz'_e | \seededge, \focalnode, \at{\vz}{t_e-1}; \paramvector) \lik(\seededge,\focalnode | \at{\edges}{t_e-1}, \at{\vz}{t_e-1}; \paramvector) \;, \label{eq:likelihood}
    \end{align}
\end{widetext}
    where $\at{\vz}{t_e-1}$ is the vector of node labels immediately before $e$ was added, and $\vz'_e$ is the vector of node labels added with edge $e$ at time $t_e$.
    The term $\lik(\seededge,\focalnode | \at{\edges}{t_e-1}, \at{\vz}{t_e-1}; \paramvector)$ gives the likelihood of selecting edge $\seededge$ and node $\focalnode$ at time $t - 1$; its value is $\frac{1}{\abs{\seededge}\at{m}{t-1}}$ if $\seededge \in \at{\edges}{t_e-1}$ and $0$ otherwise, where $\at{m}{t-1} = \abs{\at{\edges}{t_e-1}}$ is the number of edges at time $t - 1$.
    The term $\lik(\newedge, \vz'_e | \seededge, \focalnode, \at{\vz}{t_e-1}; \paramvector)$ gives the likelihood of realizing edge $\newedge$ and the associated new node labels $\vz'_e$ given that edge $\seededge$ and node $\focalnode$ were selected at time $t - 1$.
    A helpful note for computation is that $\lik(\newedge, \vz'_e | \seededge, \focalnode, \at{\vz}{t_e-1}; \atalg{\hat{\paramvector}}{\ell}) = 0$ if either $\focalnode \notin \newedge\cap \seededge$ or $\seededge\not\prec \newedge$.

    The requirement in \cref{eq:likelihood} to sum over the latent variables $f$ and $u$ at each timestep implies a computational cost to evaluate $\lik(\hypergraph, \vz ; \paramvector)$ (or its derivatives) which is quadratic in the number of edges $m$, making direct optimization of the likelihood expensive for data sets of even modest size. 
    Batch expectation-maximization \cite{dempster1977maximum}  is an alternative approach which maximizes the likelihood without explicitly computing \cref{eq:likelihood}, but even this approach requires forming a belief distribution over the values of all latent variables at each timestep. 
    This again incurs a quadratic cost per iteration. 
    
    We therefore use a stochastic expectation-maximization (SEM) algorithm \cite{cappeOnLineExpectationMaximization2009} to estimate the parameters $\paramvector$.
    Each iteration of SEM consists of two steps: the stochatic E-step, in which noisy averages of the sufficient statistics with polynomially-decaying rates are calculated using the current parameter estimates, and the stochastic M-step, in which the parameter estimates are updated using the expected sufficient statistics. 

    To start the SEM algorithm, we initialize a vector $\atalg{\sufficientstatsvec}{0}$ of sufficient statistics to arbitrary values; in the \model{} model there are two sufficient statistics per copy parameter and one sufficient statistic per each of the four addition parameters, so $\sufficientstatsvec$ has length $8$.
    From $\atalg{\sufficientstatsvec}{0}$, we compute an initial estimate of the parameter vector $\atalg{\hat{\paramvector}}{0} = g(\atalg{\sufficientstatsvec}{0})$ via a function $g$ which maps sufficient statistics to parameter estimates; details on both the sufficient statistics $\sufficientstatsvec$ and the mapping $g$ are provided in \Cref{sec:sufficient-statistics}.
    We then update a moving average of $\sufficientstatsvec$ as follows. 
    In each algorithmic step $\ell$, we pick an edge $e$ uniformly at random from $\hypergraph$. 
    Then, given current estimates $\atalg{\hat{\paramvector}}{\ell}$, we form a belief distribution over the latent identities of $\seededge \in \edges$ and $\focalnode \in \seededge$ which could have been selected in the edge-
    step to form $\newedge$. 
    Using Bayes' rule, this probability is 
    \begin{widetext}
    \begin{align}
        \lik(\seededge,\focalnode | \newedge, \vz'_e; \atalg{\hat{\paramvector}}{\ell}) = \frac{\lik(\newedge, \vz'_e | \seededge, \focalnode, \at{\vz}{t_e-1}; \atalg{\hat{\paramvector}}{\ell}) \abs{\seededge}^{-1}}{\sum_{\seededge' \prec \newedge} \sum_{\focalnode' \in \seededge'} \lik(\newedge, \vz'_e | \seededge', \focalnode', \at{\vz}{t_e-1}; \atalg{\hat{\paramvector}}{\ell}) \abs{\seededge'}^{-1} } \;. \label{eq:bayes}
    \end{align}
    \end{widetext}
    We then form the vector of expected sufficient statistics 
    \begin{align}
        s' = \sum_{\seededge \prec \newedge} \sum_{\focalnode \in \seededge\cap \newedge} \lik(\seededge, \focalnode | \newedge, \vz'_e; \atalg{\hat{\paramvector}}{\ell}) \vpsi(\newedge, \seededge, \focalnode, \vz) \;, \label{eq:sufficient-stats-update}
    \end{align}
    where $\vpsi(\newedge, \seededge, \focalnode, \vz)$ is the vector of sufficient statistics for the parameters $\paramvector$ given $\newedge$, $\seededge$, $\focalnode$, and $\vz$.
    The entries of $\vpsi(\newedge, \seededge, \focalnode, \vz)$ can be computed as functions of the node labels in $\newedge$ and $\seededge$; we supply explicit formulae in \Cref{sec:sufficient-statistics}. 
    We then perform the update 
    \begin{align*}
        \atalg{\sufficientstatsvec}{\ell+1} = (1 - \atalg{\lr}{\ell} )\atalg{\sufficientstatsvec}{\ell} + \atalg{\lr}{\ell} \sufficientstatsvec' \;,
    \end{align*}
    where $\atalg{\lr}{\ell}$ is the learning rate at iteration $\ell$; we use a polynomially decaying learning rate in our experiments.
    Our new estimate of the parameter vector $\paramvector$ is then given by a function $\atalg{\hat{\paramvector}}{\ell+1} = g(\atalg{\sufficientstatsvec}{\ell+1})$ of the sufficient statistics, which we also supply in \Cref{sec:sufficient-statistics}.
    \Cref{alg:sem} summarizes this process. 
    Our version of the algorithm uses an online compuation of $\sufficientstatsvec'$, which allows us to perform the double-sum appearing in \cref{eq:sufficient-stats-update,eq:bayes} only once, and without explicitly storing the resulting belief distribution at any point.

    \begin{algorithm}[H]
        \caption{\model{} SEM Update Step}\label{alg:sem}
        \begin{algorithmic}
        \Require $\hypergraph = \paren{\nodes, \edges}$, $\vz$, $\atalg{\hat{\paramvector}}{\ell}$, $\atalg{\sufficientstatsvec}{\ell}$
        \State $e \gets \text{Uniform}(\edges)$
        \State $\vx \gets \vzero$
        \State $\vy \gets \vzero$
        \For{$\seededge \in \edges$}
            \If{$\seededge \prec \newedge$}
                \For{$\focalnode \in \newedge \cap \seededge$}
                    \State $\vx \gets \vx + \lik(\seededge, \focalnode | \newedge; \atalg{\hat{\paramvector}}{\ell}) \vpsi(\newedge, \seededge, \focalnode, \vz)$
                    \State $\vy \gets \vy + \lik(\seededge, \focalnode | \newedge; \atalg{\hat{\paramvector}}{\ell})$
                \EndFor
            \EndIf
        \EndFor
        \State $\sufficientstatsvec' \gets \vx \oslash \vy$ \Comment{entrywise division}
        \State $\atalg{\sufficientstatsvec}{\ell+1} \gets (1 - \atalg{\lr}{\ell} )\atalg{\sufficientstatsvec}{\ell} + \atalg{\lr}{\ell} \sufficientstatsvec'$
        \State $\atalg{\hat{\paramvector}}{\ell+1} \gets g(\atalg{\sufficientstatsvec}{\ell+1})$
        \State \Return $\atalg{\hat{\paramvector}}{\ell+1}$, $\atalg{\sufficientstatsvec}{\ell+1}$
        \end{algorithmic}
    \end{algorithm}

    We perform SEM by repeating \Cref{alg:sem} until a stopping criterion is met. 
    For our criterion, we define the function 
    \begin{align*}
        h(\atalg{\hat{\paramvector}}{\ell}, \atalg{\hat{\paramvector}}{\ell'}) = \max_{i} \left\{\abs{\atalg{\hat{\param}_i}{\ell} - \atalg{\hat{\param}_i}{\ell'}} - \epsilon_\mathrm{abs} - \epsilon_\mathrm{rel} \atalg{\hat{\param}_i}{\ell'}\right\} \;.
    \end{align*}
    The function $h$ here can be viewed as combining absolute and relative error tolerances, enabling the algorithm to declare convergence even in cases when the parameter estimates are near zero. 
    In our implementation, we declare convergence at timestep $\ell$ when $h(\atalg{\hat{\paramvector}}{\ell}, \atalg{\hat{\paramvector}}{\ell'}) \leq 0$ for all $\ell'$ in the most recent 500 iterations of the algorithm, with tolerances $\epsilon_\mathrm{rel} = 0.015$ and $\epsilon_\mathrm{abs} = 0.01$.

\subsection{Community Detection via Simulated Annealing} \label{sec:community-detection}

    We also consider the community detection problem of learning the node labels $\vz$ given an observed hypergraph $\hypergraph$ and a fixed parameter vector $\paramvector$.
    Since \cref{eq:likelihood} defines a likelihood function over both the parameter vector $\paramvector$ and the node labels $\vz$, we can attempt to also learn $\vz$ via maximum-likelihood estimation.
    Unlike in stochastic block model variants in which there are several heuristics with favorable computational properties \cite{newman2016equivalence,chodrowGenerativeHypergraphClustering2021,chodrowNonbacktrackingSpectralClustering2023,ruggeriCommunityDetectionLarge2023,kritschgauCommunityDetectionHypergraphs2024,hood2026broad,li2026higher}, the edge dependence structure encoded by \model{} makes it difficult to apply standard efficient heuristics.
    The development of alternative efficient heuristics is an important direction of future work. 
    In our present approach, we therefore attempt to learn $\vz$ via simulated annealing \cite{kirkpatrickOptimizationSimulatedAnnealing1983},  a general-purpose optimization algorithm which is typically used when other heuristics are not available. 
    Our primary aim is to demonstrate that likelihood maximization in this model can detect label structure in at least some circumstances where other community detection methods perform poorly, thereby motivating the development of more efficient optimization heuristics in future work. 
    
    A direct evaluation of \cref{eq:likelihood} requires $O(m^2)$ terms, making it expensive to evaluate even for data sets of modest size. 
    An important computational limitation of simulated annealing in this context is that the label $z_i$ of node $i$ appears in \emph{all} terms of the likelihood \cref{eq:likelihood} corresponding to edges after which node $i$ arrives. 
    This includes edges in which node $i$ does not participate, implying that we must perform $O(m^2)$ computations even to evaluate the change in likelihood under a proposed change to $z_i$.
    This contrasts to, for example, modularity \cite{newman2004finding}, where the computational cost of moving a node between communities is constant with appropriate use of data structures.

    To mitigate the computational burden, we use an intersection-based approximation to the likelihood of forming edge $\newedge$ in which we only sum over candidate copied edges $\seededge$ such that $\abs{\seededge \cap \newedge}$ is large.
    The intuition of this approximation, which we justify in detail in \Cref{sec:simulated-annealing-approximation}, is that $\lik(\newedge, \vz_\newedge'|\seededge, \focalnode, \vz; \paramvector)\in O(n^{-x})$, where $n$ is the number of nodes in the hypergraph and $x = \abs{\newedge} - \abs{\newedge\cap \seededge} - \abs{\newedge \setminus \nodes}$ is the number of extant nodes required to form $\newedge$ given $\seededge$. 
    The essence of the argument is that forming $x$ extant nodes requires $x$ independent samples from the sets $\nodes_0\setminus \seededge$ and $\nodes_1\setminus \seededge$, each of which has probability $O(n^{-1})$ of being selected.
    We  therefore design our algorithm to maximize the objective function $\tilde{\cL}(\vz) = \log \tilde{\lik}(\hypergraph, \vz ; \paramvector)$
    where $\tilde{\lik}(\hypergraph, \vz ; \paramvector)$ is the likelihood of $\hypergraph$ and $\vz$ under the intersection-based approximation.
    A single epoch consists of $n$ individual update steps. 
    In an update step in epoch $\ell$, we propose a change to the labels $\vz$ by generating a node index $j$ uniformly at random and evaluating the change in likelihood associated with flipping the binary node label ($z_j \gets \bar{z}_j$).  
    If the proposal increases $\tilde{\cL}(\vz)$, it is accepted.
    If the proposal decreases $\tilde{\cL}(\vz)$, however, we may also accept it with an acceptance probability $\atalg{\acceptprob}{\ell}$ that decreases with the epoch $\epoch$ and depends on the magnitude of the decrease in $\tilde{\cL}(\vz)$.

    \begin{algorithm}[H]
        \begin{algorithmic}
            \caption{\texttt{SimulatedAnnealing}$(\hypergraph, \paramvector)$}\label{algocf:simulated_annealing}
            \Require{Hypergraph $\hypergraph$, parameters $\paramvector$, number of epochs $\maxepochs$}
            \Ensure{Label set $\hat{\vz}$, estimate of true labels $\vz$}
            \State $\sigma_\hypergraph$, $\hat{\vz}$ $\gets$ \texttt{SimulatedAnnealingFirstEpoch}$(\hypergraph, \paramvector)$
            \For{$\epoch = 1$ to $\maxepochs$}
                \For{$\step = 1$ to $n$}
                    \State Sample $j$ $\sim$ Uniform($\{1,...,n\}$)
                    \State $\hat{\vz}^\prime$ $\gets$ $\hat{\vz}$
                    \State $z^\prime_j\gets1-z^\prime_j$
                    \State $\delta \gets (\log \tilde{\lik}(\hypergraph, \hat{\vz}^\prime ; \paramvector) - \log \tilde{\lik}(\hypergraph, \hat{\vz} ; \paramvector))/\sigma_\hypergraph$
                    \If{$\delta>0$}
                        \State $\hat{\vz} \gets\hat{\vz}^\prime$
                    \Else
                        \If{$2 \leq \epoch \leq 5$}
                            \State $\atalg{\acceptprob}{\epoch}$ $\gets$ $\mathrm{Gaussian}(\delta;\mu = 0, \sigma=2)$
                        \ElsIf{6 $\leq \epoch$}
                            \State $\atalg{\acceptprob}{\epoch}$ $\gets$ $\mathrm{Gaussian}(\delta;\mu = 0, \sigma=2(1 - \frac{\epoch}{\maxepochs}))$
                        \EndIf
                        \State Sample $x \sim \text{Uniform}(0,1)$
                        \If{$x < \atalg{\acceptprob}{\epoch}$} 
                            \State $\hat{\vz} \gets \hat{\vz}^\prime$
                        \EndIf
                    \EndIf
                \EndFor
            \EndFor
            \State \Return$\hat{\vz}$
        \end{algorithmic}
    \end{algorithm}

    We also use a custom acceptance schedule $\atalg{\acceptprob}{\ell}$ for our simulated annealing implementation.
    Our approach is motivated by the observation that, in our application, the change in the objective $\tilde{\cL}$ of a proposal can vary significantly depending on the size of the hypergraph under study.
    Thus, in our early experiments, traditional annealing schedules such as harmonic or exponential decay struggled to still occasionally accept proposals that decrease $\tilde{\cL}$. 
    Our custom implementation instead takes an adaptive approach based on the distribution of proposals previously observed. 
    In the first epoch, we accept \emph{all} proposals, while storing the changes to the objective $\tilde{\cL}$ produced by those proposals. 
    At the end of the epoch, we compute the standard deviation $\sigma(\Delta \tilde{\cL})$ of the changes in $\tilde{\cL}$ observed during the epoch. 
    This procedure is handled by the function \texttt{SimulatedAnnealingFirstEpoch} in our implementation (\Cref{algocf:simulated_annealing}).
    For the remaining epochs, we accept proposals with a probability obtained by evaluating a Gaussian density with mean 0 and a standard deviation proportional to $\sigma(\Delta \tilde{\cL})$, with constant of proportionality that decays with epoch $\ell$.
    We provide a detailed discussion of this algorithm, including our approximate objective function, the implementation of the annealing schedule, and the first epoch of the algorithm, in \Cref{sec:simulated-annealing-algorithm-spec}.

\section{Results} \label{sec:results}

    We now test the ability of our proposed \model{} model to recover parameters (via SEM) and node labels (via simulated annealing) on both synthetic and empirical data sets.

\subsection{Parameter Estimation on Synthetic Data}

\begin{figure*}[t]
    \includegraphics[width=\textwidth]{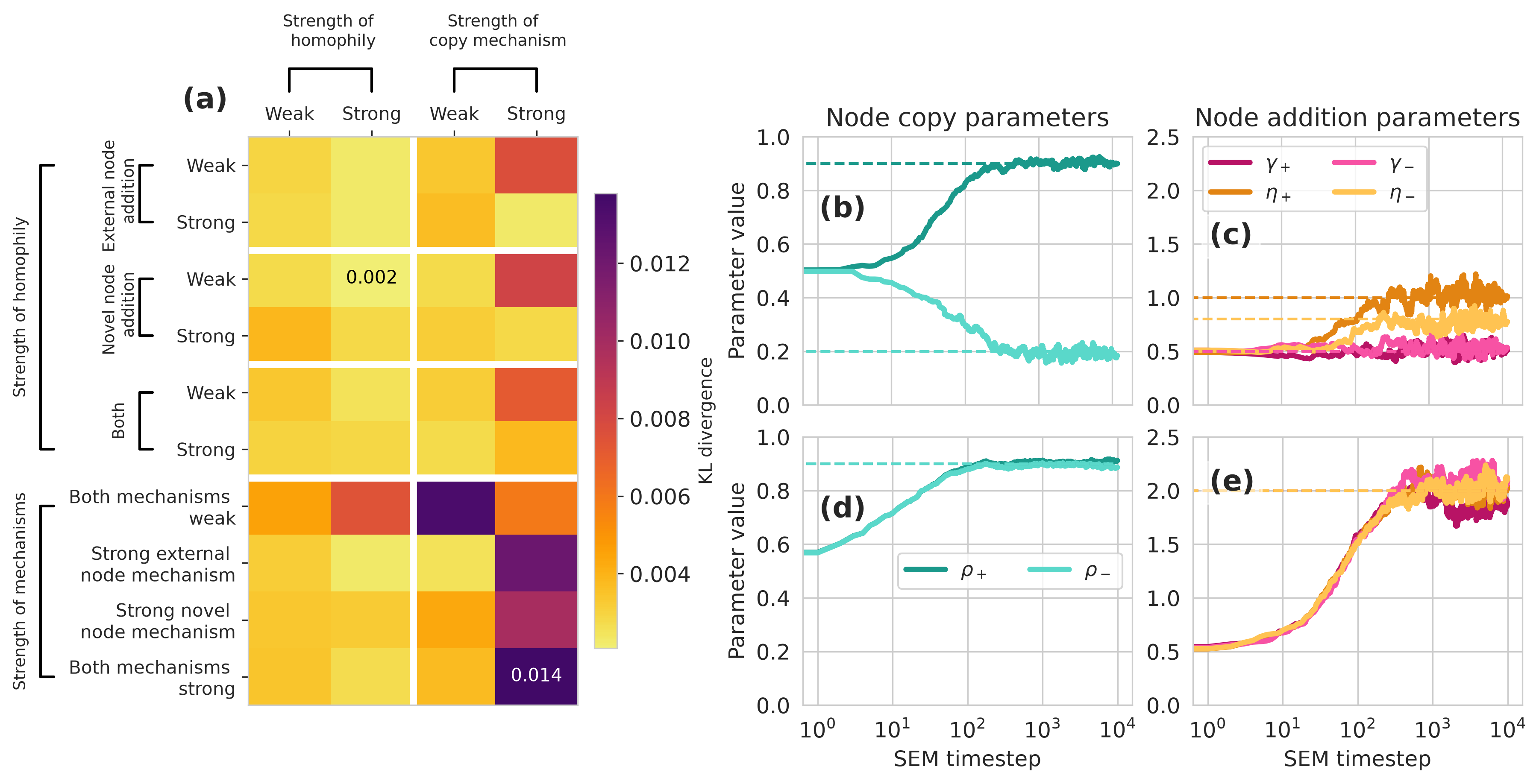}
    \caption{
        Performance of SEM on synthetic hypergraphs. 
        Each pixel in the heatmap (panel (a)) corresponds to a set of parameters $\paramvector$, which can be found in Appendix \ref{sec:SEM-appendix}.        
        For each set of parameters, 20 independent hypergraphs are simulated under the \model{} model for 2,000 simulation steps. 
        We then perform SEM on each.  
        Finally, we compute the Kullback-Leibler (KL) divergence of the estimated parameters from the true ones (\cref{eq:kl-divergence}) and average the results to obtain a value in a pixel. 
        The parameter estimates over iterations of SEM are plotted for the underlying parameters which produced the lowest mean KL divergence (copy parameters in panel (b) and addition parameters in panel (c)), 
        and similarly for the parameters that produced the highest mean KL divergence in panels (d) and (e).}
    \label{fig:synthetic_sem}
\end{figure*}

\Cref{fig:synthetic_sem} illustrates the behavior of SEM in recovering parameters from synthetic hypergraphs generated by the \model{} model.
Each synthetic hypergraph is generated by simulating \model{} with a known parameter vector $\paramvector$.
We say that an instance of the \model{} model possesses \textit{homophily} in the edge copy step of the growth process if $\copyrate_+ > \copyrate_-$, and \emph{heterophily} if $\copyrate_+ < \copyrate_-$. 
We refer to homophily and heterophily in the extant and novel node addition steps in parallel fashion.
We perform SEM on a set of hypergraphs that were generated with varying strengths of homophily across each of the three growth mechanisms: the edge copy step, the extant node addition step, and the novel node addition step.
We also vary the expected number of nodes added to $e$ in each of these steps.
To measure the success of a parameter recovery, we compute the Kullback-Leibler (KL) divergence between the true and estimated distributions described by each of the six parameters, and add the results. 
Using standard results, our measure of error is given by 
\begin{align}
    \delta_{\mathrm{KL}} &= \phantom{+}\sum_{* \in \{+, -\}} c_*\left( \copyrate_{*} \log \frac{\copyrate_{*}}{\hat{\copyrate}_{*}} +  (1 - \copyrate_{*}) \log \frac{1 - \copyrate_{*}}{1 - \hat{\copyrate}_{*}} \right) \nonumber\\ 
    &\phantom{=} + \sum_{* \in \{+, -\}} \left( \extantrate_{*} \log \frac{\extantrate_{*}}{\hat{\extantrate}_{*}} - \extantrate_* + \hat{\extantrate}_{*} \right) \label{eq:kl-divergence} \\ 
    &\phantom{=} + \sum_{* \in \{+, -\}} \left( \novelrate_{*} \log \frac{\novelrate_{*}}{\hat{\novelrate}_{*}} - \novelrate_* + \hat{\novelrate}_{*} \right)\;, \nonumber
\end{align}
where $c$ is a factor describing, in expectation, how many nodes of each label are selected in the copy step. 
For this analysis, we set $c_* = 1$ for both labels.

As seen in \Cref{fig:synthetic_sem}, the quality of the learned parameter estimates depends on the seeded parameter vector. 
We observe especially low estimation errors when the seeded parameters reflect strong homophily in the edge copy step as reflected in the difference $\same{\copyrate} - \opp{\copyrate}$; a typical SEM run for these parameter settings is shown in the top row on the right of \Cref{fig:synthetic_sem}.
That said, even in the least successful cases for estimation with high KL divergence, the learned estimates are still visually close to the true values, illustrated in the bottom row on the right of \Cref{fig:synthetic_sem}.

\begin{table*}
    \caption{Data sets used in this paper. We show the number of nodes $n$, edges $m$, average degree $\langle d \rangle$, and average edge size $\langle k \rangle$ for each data set.
    All data sets are used for parameter estimation. Only \texttt{senate-bills}, \texttt{high-school}, and \texttt{primary-school} are used in community detection experiments.}\label{tab:datasets}
    \begin{tabular}{lrrrrrrr}
        & $n$ & $m$ & $\langle d \rangle$ & $\langle k \rangle$ & Nodes are & Edges are & Labels represent\\
        \midrule
        \texttt{senate-bills} \cite{fowler2006connecting,fowler2006legislative,chodrowGenerativeHypergraphClustering2021} & 294 & 29,157 & 789.6 & 8.0 & Senators & Cosponsored bills & Political party\\
        \texttt{house-bills} \cite{fowler2006connecting,fowler2006legislative,chodrowGenerativeHypergraphClustering2021} &  1,494 & 60,987 & 835.8 & 20.5 & Congresspeople & Cosponsored bills & Political party  \\
        \texttt{high-school} \cite{gemmetto2014mitigation,stehle2011high} & 320 & 60,544 & 484.6 & 2.6 & Students & Social interactions & Gender  \\
        \texttt{primary-school} \cite{mastrandrea2015contact} & 227 & 44,626 & 590.7 & 3.0 & Students & Social interactions & Gender \\
        \texttt{coauthorship} \cite{agarwal2016women} & 105,256 & 142,834 & 3.4 & 2.5 & Authors & Papers & Gender \\
        \texttt{enron-emails} \cite{klimt2004enron}  & 14,901 & 83,372 & 18.5 & 3.3 & Email addresses & Email threads & Fringe vs. core \\
        \bottomrule
    \end{tabular}
\end{table*}

\subsection{Parameter Estimation on Empirical Data} \label{sec:sem-empirical}

We also used our method to infer parameters for six empirical hypergraph data sets described in \Cref{tab:datasets}. 
Although many labeled hypergraph data sets are now available for community detection, relatively few have natively binary node labels. 
The \texttt{senate-bills} and \texttt{house-bills} data sets \cite{fowler2006connecting,fowler2006legislative} are hypergraphs of cosponsorships of bills in the U.S. Senate and House of Representatives, respectively, with node labels corresponding to the political party of each legislator; we use the version of these data sets used in \cite{chodrowGenerativeHypergraphClustering2021}.
The \texttt{high-school} and \texttt{primary-school} data sets \cite{gemmetto2014mitigation,stehle2011high,mastrandrea2015contact} are hypergraphs of social interactions between students in a high school and primary school, respectively, with node labels corresponding to the gender of each student; these data sets were collected through the SocioPatterns project.
In these data sets, we consider a sequence of interactions to correspond to a single hyperedge if those interactions are identical as sets and temporally contiguous. 
The \texttt{coauthorship} data set \cite{agarwal2016women} contains publishing records for 81 computer science conferences from 2000 to 2015, with an inferred binary gender label for each author.
Finally, we considered the \texttt{enron-emails} data set, which contains the email records for 145 users associated with the Enron Corporation \cite{klimt2004enron}.

\begin{figure}[t]
    \includegraphics[width=0.5\textwidth]{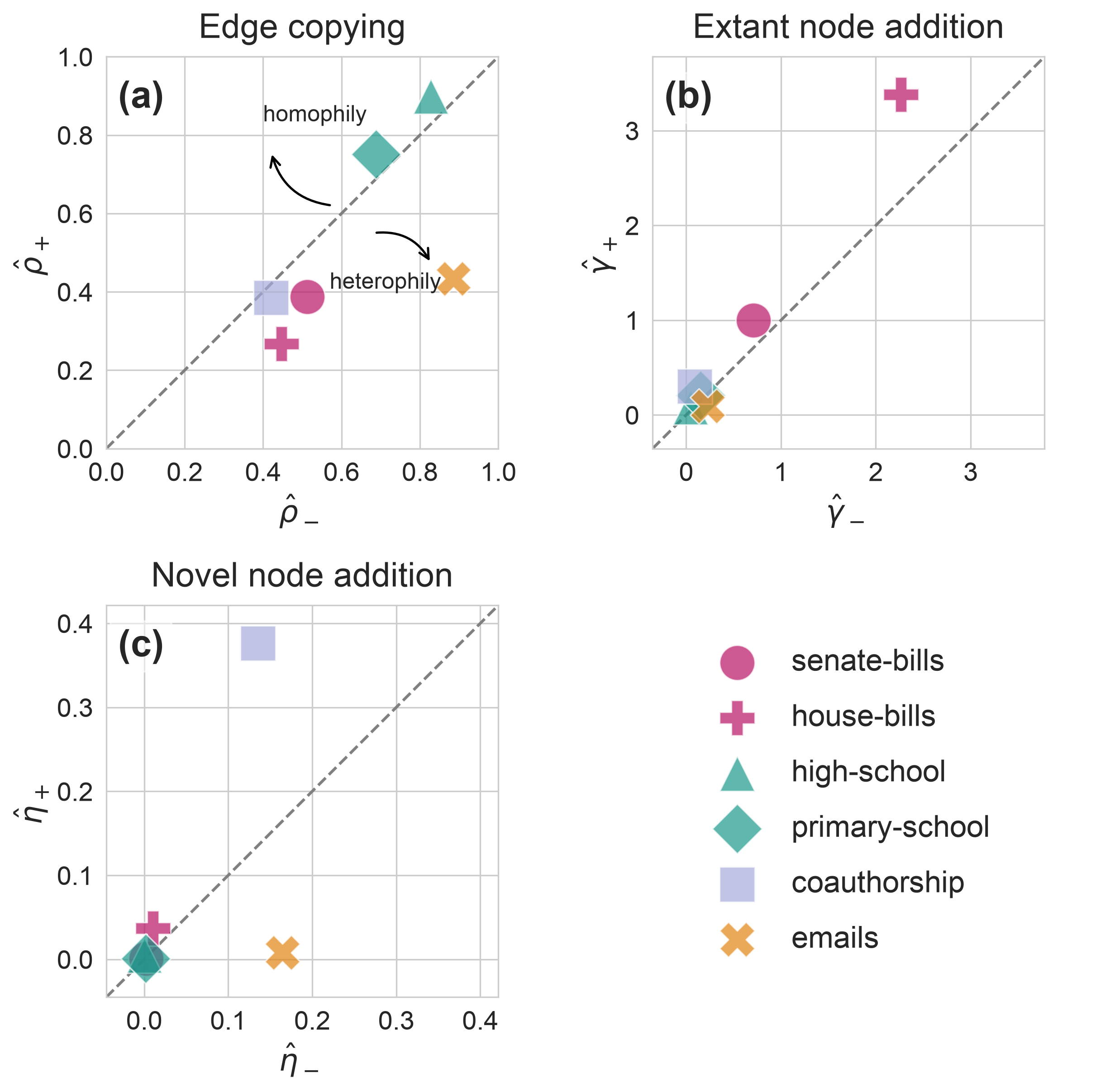}
    \caption{
        Model parameters inferred via SEM for six empirical higher-order networks.
        The line of equality is shown as a guide for the eye.
        }
     \label{fig:empirical_sem}
\end{figure}

\Cref{fig:empirical_sem} shows our parameter estimates for each of the empirical data sets under the \model{} model. 
In each panel, we plot the same-label and opposite-label parameter estimates on opposing axes. 
Some data sets, like the gendered \texttt{primary-school} contact network, appear on or near the line of equality in all three panels, suggesting that homophily with respect to the given binary labels is not a major driver of network evolution. 
The \texttt{coauthorship} data, meanwhile, falls on the line of equality for the edge copy and extant node addition mechanisms, but there is a difference in the novel node addition parameters: $\same{\hat{\novelrate}} > \opp{\hat{\novelrate}}$.
Our model presents a simplified picture of the process of collaboration formation as being initiated by a focal individual scientist, who then recruits from a previous collaboration, from other published scientists in the field, and from an unobserved pool of novices who have never published before. 
Our findings suggest that of these mechanisms, only the latter (the recruitment of novices) displays substantial gender homophily, with novices who share the gender of the focal scientist more likely to be recruited to the collaboration. 
This effect may be partially due to gender homophily between PhD students and their advisors, a known phenomenon in several scientific areas \cite{schwartz2022impact}. 

Both the \texttt{senate-bills} and \texttt{house-bills} data sets appear to have weak homophily in the estimates of the extant node addition parameters and weak heterophily in the edge copy parameters.
Because new congresspeople are elected far less frequently than bills are cosponsored, the novel node addition parameters are nearly 0 for both data sets. 
The slight heterophily in the copy step of \texttt{senate-bills} and \texttt{house-bills} may be related to reciprocity in bipartisan cosponsorship, a phenomenon that has been observed in U.S. legislative bodies \cite{harbridge2023bipartisan,dobson2026selective}.
We illustrate the connection between reciprocity and edge copy parameters through the following example.
Consider some bill $\newedge$, where $\focalnode$ is a cosponsor of both bill $\newedge$ and bill $\seededge$.
If $\focalnode$ is a Democrat, for instance, and the primary sponsor of $\seededge$ is a Republican, then $\focalnode$ was a ``bipartisan'' cosponsor of $\seededge$.
The reciprocity of bipartisan cosponsorship suggests that the Republican cosponsors of $\seededge$ may also cosponsor $\newedge$ to return the favor.
However, the other Democratic cosponsors of $\seededge$ are not indebted in the same way to $\focalnode$, possibly making them less likely to cosponsor $\newedge.$
Further investigation of the relationship between reciprocity and edge copying in legislative cosponsorship is an interesting direction for future work.

The novel node estimates for both the congressional datasets are very small. 
The largest difference in our estimates for these two systems is in the extant node addition parameter estimates; the \texttt{house-bills} data set has estimates $\same{\hat{\extantrate}}$ and $\opp{\hat{\extantrate}}$ which are each roughly three times larger than their values in the \texttt{senate-bills} data set.
These estimates suggest the operation of similar levels of party homophily in the cosponsorship of bills via extant node addition, while reflecting the higher mean edge size of \texttt{house-bills}, where a bill is co-sponsored by a mean of over 20 congresspersons, as opposed to the mean of 8 in \texttt{senate-bills}. 

The \texttt{enron-emails} hypergraph falls in the heterophilic region of the edge copy panel.
This is a feature of the sampling strategy.
The Enron data set contains the email data from 145 users associated with the Enron Corporation.
These users are the ``core'' nodes of the hypergraph.
Any users who were included in \texttt{enron-emails} with core users, but whose complete email data was not sampled, are referred to as ``fringe'' nodes.
Each hyperedge represents an email between a set of nodes, or users.
Therefore, each hyperedge must contain at least one core node.
Since 73\% of edges in the hypergraph have size 2, and less than one percent of the nodes in the hypergraph are core nodes, most edges (67\%) join a core node to a fringe node, generating disassortativity with respect to these labels.  
The high magnitude of the copy parameters, particularly $\copyrate_-$, is a result of the nature of email — rarely does an email interaction between a group of users occur only once.
Often, an email is precipitated by an existing conversation between that set of users, making the edge copy mechanism well suited for modeling email network growth.

\subsection{Community Detection via Simulated Annealing}

\begin{figure}
    \includegraphics[width=\columnwidth]{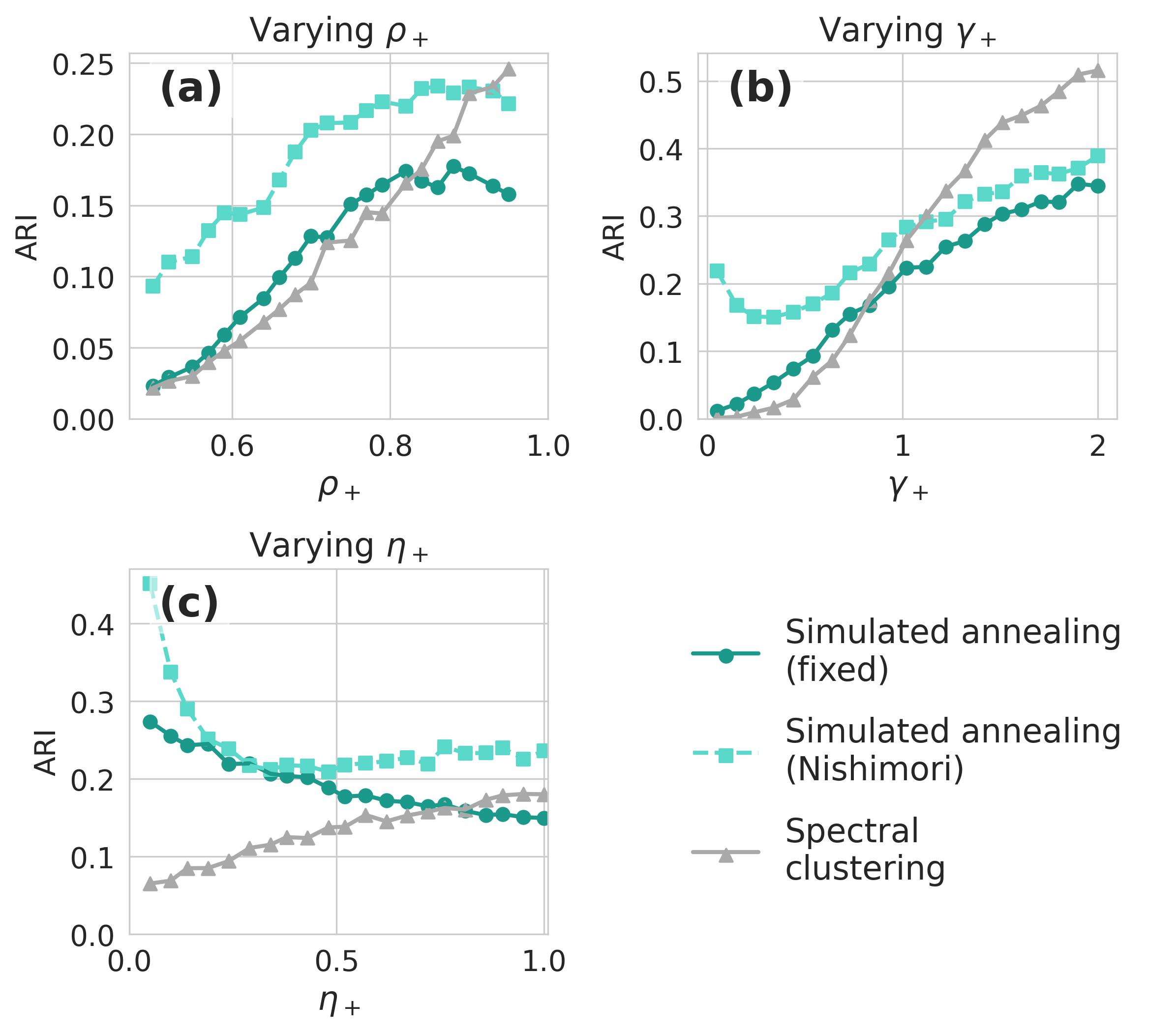}
    \caption{
        Comparison of community detection methods on synthetic data. 
        We compare our simulated annealing algorithm under the Nishimori condition and with a fixed parameter vector $\paramvector' = \thetaguess$ against a baseline of Laplacian spectral clustering on the weighted clique projection of the hypergraph. 
        The ``default'' parameter vector used in synthetic hypergraph generation is $\paramvector = (\same{\copyrate}, \opp{\copyrate}, \same{\extantrate}, \opp{\extantrate}, \same{\novelrate}, \opp{\novelrate}) = (\synthfixedetaplus,\synthfixedetaminus, \synthfixedlambdaplus, \synthfixedlambdaminus,\synthfixedgammaplus, \synthfixedgammaminus)$. 
        In each of the three panels (a-c), a single one of these parameters is varied while the others are held fixed at their default values. 
        Each point gives the mean ARI over 500 independent model simulations, each of which was simulated for 200 timesteps. 
    } \label{fig:community-synthetic}
\end{figure}

We evaluate our simulated annealing algorithm on a range of small synthetic data sets under two conditions, shown in \Cref{fig:community-synthetic}. 
Under the \emph{Nishimori condition} \cite{decelle2011asymptotic}, we assume that we have access to the exact parameter vector used to generate the synthetic data. 
In this condition, we therefore generate a synthetic hypergraph according to a parameter vector $\paramvector$ and then attempt to learn $\vz$ via simulated annealing using the same parameter vector $\paramvector$. 
Under the ``fixed'' condition, which simulates a more realistic data analysis scenario, we always use the same parameter vector $\paramvector'$ to learn $\vz$ via simulated annealing regardless of the true parameter vector $\paramvector$ used to generate the synthetic data. 
In this condition, we use a default value of $\paramvector' = \thetaguess$. 
We compare these methods against each other and against a baseline of Laplacian spectral clustering \cite{luxburg2007tutorial} applied to the weighted clique projection of the hypergraph. 
We assess the effectiveness of these techniques and our method with a measure of clustering effectiveness, the adjusted Rand index (ARI) \cite{hubert1985comparing}.
In each panel of \Cref{fig:community-synthetic}, we show the performance of each of these models as a single entry of $\paramvector$ is varied while the others are held fixed. 
Unsurprisingly, simulated annealing under the Nishimori condition outperforms simulated annealing under the fixed condition. 
For strong homophily (larger values of the varied parameter), simulated annealing under the Nishimori condition can nevertheless be outperformed by Laplacian spectral clustering. 
Notably, however, there are regions of parameter space, especially those with weak homophily (smaller values of the varied parameter), in which both simulated annealing under the fixed condition and under the Nishimori condition outperform Laplacian spectral clustering. 
This result is in some sense unsurprising, since we perform our experiment on synthetic samples drawn directly from the \model{} model.
Nevertheless, these experiments suggest that there exist classes of hypergraphs in which community structure can be more reliably detected via likelihood maximization under \model{} than by at least this standard graph method. 

Indeed, we observe relatively favorable performance of the \model{} model on empirical data as well. 
We evaluate our simulated annealing algorithm on a subset of our empirical data sets: \texttt{senate-bills}, \texttt{primary-school}, and \texttt{high-school}; the poor scaling of simulated annealing with the number of nodes $n$ made it computationally prohibitive to run on the larger \texttt{house-bills}, \texttt{enron-emails}, and \texttt{coauthorship} data sets.
We use the class membership of the students in \texttt{primary-school} and \texttt{high-school} as the ground truth labels in this section instead of gender because community structure has been previously detected with these labels \cite{chodrowGenerativeHypergraphClustering2021}.  
For our experiments we compare the binary labels generated by our simulated annealing algorithm against the multinary ground-truth class labels. 
Since in the context of community detection the labels $\vz$ (and therefore parameter estimates $\hat{\paramvector}$) are not available, we instead use a single, fixed parameter vector for each of these three experiments: $\hat{\paramvector} = (0.90, 0.10, 1.00, 0.25, 0.001, 0.001)$.
The zero values for the novel node addition parameters reflect a choice of algorithmic convenience. 
All of these data sets are appropriately modeled with very small values of $\same{\novelrate}$ and $\opp{\novelrate}$ under our model due to their large number of edges compared to the small number of nodes.  
To simplify computation, we instead simply assume that all nodes exist in the system at all times, even if they have not yet participated in an edge. 

We compare \model{} simulated annealing against several competitors. 
The all-or-nothing hypergraph modularity of \cite{chodrowGenerativeHypergraphClustering2021} does not allow the user to specify a desired number of clusters, but because it returns a binary partition on \texttt{senate-bills}, we can compare directly on this data set. 
Belief-propagation nonbacktracking spectral clustering \cite{chodrowNonbacktrackingSpectralClustering2023} allows the user to specify the number of clusters, and the authors report results for binary clustering. 
We therefore compare our method against the reported results from these two prior works.\footnote{The authors of \cite{chodrowNonbacktrackingSpectralClustering2023} did not perform the data filtering on \texttt{high-school} and \texttt{primary-school} described in \Cref{sec:sem-empirical}, so our direct comparison with their results should be interpreted with some caution.}
Notably, both hypergraph modularity maximization and nonbacktracking spectral clustering are based on generative models which assume conditional edge independence, the assumption relaxed by \model{}.
We also include as comparisons two dyadic methods, implemented on the weighted clique projection of the hypergraph, that allow the user to specify the number of clusters: Laplacian spectral clustering \cite{luxburg2007tutorial} and greedy modularity maximization \cite{clauset2004finding}, both of which also have connections to generative models that assume conditional (dyadic) edge independence \cite{newman2016equivalence,lei2015consistency}.

\begin{figure}
    \includegraphics[width=\columnwidth]{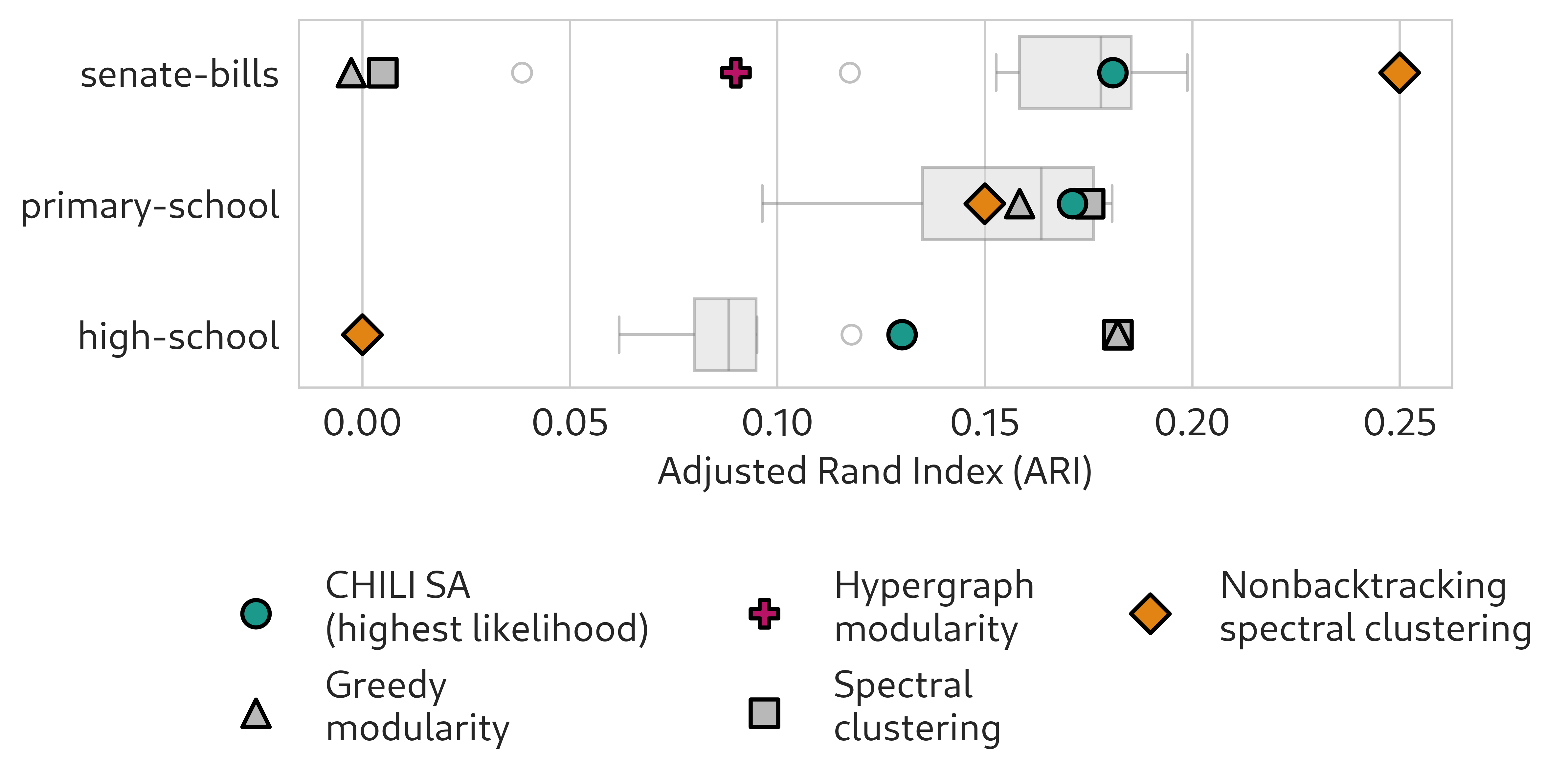}
    \caption{Community detection results on three selected empirical data sets.  
    Box plots describe the distribution of the ARI of \model{} simulated annealing (SA) on the selected data set across 10 separate runs. 
    Green filled circles give the ARI achieved by the single one of these runs to achieve the highest likelihood under simulated annealing.
    We compare these results with the ARI of the labels generated via several comparisons: hypergraph modularity \cite{chodrowGenerativeHypergraphClustering2021} (only for \texttt{senate-bills}), belief-propagation nonbacktracking spectral clustering \cite{chodrowNonbacktrackingSpectralClustering2023}, greedy dyadic modularity maximization \cite{clauset2004finding}, and dyadic Laplacian spectral clustering \cite{luxburg2007tutorial}; the latter two dyadic methods are denoted by hollow markers.  
    Values for hypergraph modularity maximization and nonbacktracking spectral clustering are reported in \cite{chodrowGenerativeHypergraphClustering2021} and \cite{chodrowNonbacktrackingSpectralClustering2023}, respectively.}
    \label{fig:community-empirical}
\end{figure}

In \Cref{fig:community-empirical}, we observe that \model{} simulated annealing outperforms both of the dyadic methods---greedy-modularity and spectral clustering---on \texttt{senate-bills} in all 10 runs in ARI.  
On \texttt{primary-school}, simulated annealing performs comparably in ARI to the competitor methods, with a slightly lower average ARI compared to spectral clustering and greedy modularity's results.  
However, simulated annealing performs worse on \texttt{high-school} in ARI compared to its competitors.
The relatively consistent performance of \model{} simulated annealing across these three data sets suggests that explicitly modeling dependence between edges may provide promising future directions in community detection, motivating a need for much more efficient heuristics with which to carry out the optimization.


\section{Discussion} \label{sec:discussion}

    We have proposed \Model{} (\model{}), a generative model for the formation of binary-labeled hypergraphs in which edges form as noisy copies of previous edges to which extant and novel nodes are also added.
    The copying, extant, and novel node addition steps are all tuned by node labels and parameters which control the strength of homophily in each of these steps.
    We have demonstrated efficient parameter inference in the \model{} model via stochastic expectation maximization (SEM). 
    We have also demonstrated that, in some circumstances, maximum-likelihood inference of node labels $\vz$ via simulated annealing, though computationally expensive, can outperform both dyadic and recent polyadic community detection algorithms that proceed from the assumption of conditional edge independence. 
    We interpret this latter result as proof-of-concept that explicitly modeling edge-dependence can be a useful framework for community detection in hypergraphs. 
    
    Considering the considerable theoretical and computational complications to modeling edge-dependence in hypergraphs, an important question is: \emph{when is it worth the effort to model edge-dependence?} 
    The answer to this question hinges on the analytic aims and the data under analysis.
    It is tempting to suppose that, as the number of nodes and edges grows large, the effects of edge-dependence will become asymptotically negligible and that simpler models may be preferred. 
    The situation, however, is somewhat unclear. 
    For example, many large empirical hypergraphs have very high rates of large intersections between edges; furthermore, these rates decay slowly as edges are added to the system \cite{heEdgeCorrelationsLink2025}.
    These interactions are signatures of edge-dependence operating on long-timescales, suggesting that models that factor in edge dependence may still be preferred in this setting. 
    A more systematic investigation into when explicitly modeling edge dependence supports data science with hypergraphs would be a very welcome direction of future work.  

    For some empirical hypergraphs, homophily may vary among the edges in a hypergraph, depending on their size \cite{veldt2023combinatorial}.
    Existing approaches to explicitly accounting for this dependence in growing hypergraph models \cite{laber2025effects} are not well-suited for our parameterization, where all parameters determine both the expected edge size and the amount of homophily.
    
    Another direction of future work concerns practical community detection using the \model{} model.
    Most obviously, although our results show that maximum-likelihood inference in the \model{} model is promising for detecting certain kinds of community structure, the simulated annealing method we have used is computationally impractical on data sets of even relatively modest size. 
    Developing more efficient heuristics for maximizing the likelihood over $\vz$ will be an important direction for scalable application of the \model{} model for community detection. 
    Scalability may also imply opportunities for improved performance.
    We note, for example,  that belief-propagation nonbacktracking spectral clustering \cite{chodrowNonbacktrackingSpectralClustering2023} achieves an ARI of approximately 0.25 on the \texttt{senate-bills} data set (\Cref{fig:community-empirical}), higher than \model{}. 
    That algorithm requires alternating between parameter estimation and community detection phases. 
    In principle, we could attempt a similar alternating procedure between SEM and simulated annealing in the \model{} model, but the computational cost of simulated annealing makes this approach infeasible in practice.
    Future work to develop more efficient heuristics for community detection under our model could, in principle, enable such an alternating algorithm, with potentially improved performance. 

    There are also opportunities to extend \model{} to more realistic settings and more diverse data sets. 
    One obvious extension is to multinary node labels, which would enable multiway community detection algorithms. 
    Another approach is to relax the assumption that \emph{every} edge is formed through the edge-copying mechanism; one could instead allow some edges to form spontaneously without copying an earlier edge. 
    A final possible extension is to incorporate recency-bias \cite{bensonSequencesSets2018} into the copy mechanism, so that more recent edges are more likely to be copied while edges that are sufficiently old may cease to be copied at all. 
    We hope that \model{} can provide a useful jumping-off point for future work in generative modeling and algorithms for hypergraphs with edge dependence mediated by node labels.

\begin{acknowledgments}
    PSC acknowledges support from the National Science Foundation under DMS-2407058.
    Compute was provided by Middlebury College through support from the National Science Foundation under award 1827373. 
\end{acknowledgments}

\section*{Data Availability}

    The code used to generate the results in this paper is available at \url{https://github.com/Violet-Ross/labeled-growth}.
    Data sets are publicly available through the papers cited in \Cref{tab:datasets}.

\section*{Use of Generative Language Models}

    Generative language models were used to assist the authors in implementing the model, methods and experiments shown in this paper. 
    The authors carefully reviewed and revised all code and associated outputs. 
    The authors are fully responsible for the correctness of all theory, methods and results presented in this paper.
    
\bibliographystyle{apsrev4-2}
\bibliography{refs} 

\appendix

\section{Power-Law Degree Distribution} \label{sec:degree-distribution-appendix}

    We now derive the asymptotic power law degree distribution described in \Cref{sec:degree-distribution}. 
    Our derivation follows that of He et al. \cite{heEdgeCorrelationsLink2025}, which in turn is based on a derivation by Mitzenmacher \cite{mitzenmacherBriefHistoryGenerative2004}.

    At stationarity, the mean degree $\brackets{d}$ of a node satisfies 
    \begin{align}
        \brackets{d} = \frac{m}{n}\brackets{k} \;,
    \end{align}
    where $\brackets{k} = \brackets{k_0} + \brackets{k_1}$ is the mean edge size (computable from the stationary distribution $\jointdist \paren{k_0,k_1}$), $m$ is the number of edges, and $n$ is the number of nodes.
    In expectation, there are $\novelrate{} = \opp{\novelrate{}} + \same{\novelrate{}}$ new nodes added per new edge, which means that $\frac{m}{n} \to \frac{1}{\novelrate{}}$ as $m, n \to \infty$ in expectation.
    This implies the asymptotic relation
    \begin{align}
        \brackets{d} = \frac{\brackets{k}}{\novelrate{}} \;. \label{eq:average-degree}
    \end{align}

    Define the moments 
    \begin{align}
        \moment_{ij} &= \brackets{\frac{k_i k_j}{k}} \;. 
    \end{align}
    We will first calculate the expected number of nodes of label 0 sampled in the edge-sampling step.
    Suppose that an edge $f$ is sampled with $k_0$ nodes of label 0 and $k_1$ nodes of label 1, for a total of $k = k_0 + k_1$ total nodes. 
    Conditional on this event, the probability that a node of label 0 is selected as the seed node $u$ is $\frac{k_0}{k}$, and when this occurs, the expected number of nodes of label 0 in $e$ is $1 + \same{\copyrate{}} (k_0 - 1)$. 
    If, on the other hand, a node of label 1 is selected as the seed node $u$, which occurs with probability $\frac{k_1}{k}$, then the expected number of nodes of label 0 in $e$ is $\opp{\copyrate{}} k_0$. 
    So, the expected number of nodes of label 0 copied into the new edge $e$ from $f$ is 
    \begin{align}
        \expectedcopiednodes{}_0 &= \brackets{\frac{k_0}{k}\paren{1 + \same{\copyrate{}}(k_0 - 1)} + \frac{k_1}{k} \opp{\copyrate{}} k_0} \\ 
        &= \brackets{\frac{k_0}{k}} + \same{\copyrate{}} \squarebrackets{\moment_{00} - \brackets{\frac{k_0}{k}}} + \opp{\copyrate{}} \moment_{01}\;. 
    \end{align}
    Similarly, the expected number of nodes of label 1 copied into the new edge $e$ from $f$ is 
    \begin{align}
        \expectedcopiednodes{}_1 = \brackets{\frac{k_1}{k}} + \same{\copyrate{}} \squarebrackets{\moment_{11} - \brackets{\frac{k_1}{k}}} + \opp{\copyrate{}} \moment_{01}\;. 
    \end{align}
    The expected total number of nodes copied into the new edge $e$ from $f$ is then
    \begin{align}
        \expectedcopiednodes{} = \expectedcopiednodes{}_0 + \expectedcopiednodes{}_1 = 1 + \same{\copyrate{}}(\moment_{00} + \moment_{11} - 1) + 2\opp{\copyrate{}} \moment_{01}\;.
    \end{align}
    Importantly, $\expectedcopiednodes{}$ depends only on the parameters and the moments of the joint distribution $\jointdist \paren{k_0,k_1}$, which can be computed via linear algebra as described in \Cref{sec:joint-distribution}.
    
    To proceed, we make a mean-field assumption that the degree $d$ of a node is independent of the ratio $\frac{k_0}{k}$ of label-0 nodes in the edge $\seededge$ sampled to form the new edge $\newedge$.
    Under this assumption, the probability that a single node selected in the sampling step has degree $d$ is $d\at{r}{t}(d)$, where $\at{r}{t}(d)$ is the fraction of nodes with degree $d$ at time $t$; the mean-field assumption implies that the node participates in $d$ edges, independently of the ratio $\frac{k_0}{k}$ of label-0 nodes in the edge $\seededge$ sampled to form the new edge $\newedge$.
    The expected number of nodes of degree $d$ sampled in the edge-sampling step is then $\expectedcopiednodes{} \frac{d \at{r}{t}(d)}{\brackets{d}}$, and the expected \emph{change} in the count of degree $d$ nodes through the edge-sampling step is 
    \begin{align}
        \frac{\expectedcopiednodes{}}{\brackets{d}} \squarebrackets{(d-1)\at{r}{t}(d-1) - d \at{r}{t}(d)}
    \end{align}
    Similarly, the expected number of nodes that are sampled in the extant node addition step is $\extantrate{} = \same{\extantrate{}} + \opp{\extantrate{}}$. 
    Since this step does not involve any selection proportional to degree, the expected change in the count of degree $d$ nodes through this step is
    \begin{widetext}
        \begin{align}
            \at{n}{t+1}\at{r}{t+1}(d) - \at{n}{t}\at{r}{t}(d) = \frac{\expectedcopiednodes{}}{\brackets{d}} \squarebrackets{(d-1)\at{r}{t}(d-1) - d \at{r}{t}(d)} + \extantrate{} \squarebrackets{\at{r}{t}(d-1) - \at{r}{t}(d)} \;.
        \end{align}
    \end{widetext}
    We know that at stationarity, $\at{n}{t+1} - \at{n}{t}$ is in expectation equal to $\novelrate{} = \same{\novelrate{}} + \opp{\novelrate{}}$, the expected number of novel nodes, while at stationarity $\at{r}{t+1}(d) = \at{r}{t}(d) = r(d)$ independent of $t$. 
    Thus, we obtain 
        \begin{align}
            r(d) \novelrate{}  = \frac{\expectedcopiednodes{}}{\brackets{d}} \squarebrackets{(d-1)r(d-1) - d r(d)} + \extantrate{} \squarebrackets{r(d-1) - r(d)} \;.
        \end{align}
    Rearranging gives
    \begin{align}
        \frac{r(d)}{r(d-1)} &=  \frac{\frac{\expectedcopiednodes{}}{\brackets{d}}(d-1) + \extantrate{}}{\frac{\expectedcopiednodes{}}{\brackets{d}} d + \novelrate{} + \extantrate{}} \\ 
        &= 1 - \frac{\frac{\expectedcopiednodes{}}{\brackets{d}} + \novelrate{}}{\frac{\expectedcopiednodes{}}{\brackets{d}} d + \novelrate{} + \extantrate{}} \;.
    \end{align}
    Allowing $d$ to become large (which describes the tail of the degree distribution), we have
    \begin{align}
        \frac{r(d)}{r(d-1)} &\approx 1 - \frac{1}{d} \frac{\frac{\expectedcopiednodes{}}{\brackets{d}} + \novelrate{}}{\frac{\expectedcopiednodes{}}{\brackets{d}}} \\ 
        &= 1 - \frac{1}{d} \paren{1 + \frac{\novelrate{}}{\expectedcopiednodes{} / \brackets{d}}}  \\ 
        &= 1 - \frac{1}{d} \paren{1 + \frac{\brackets{k}}{\expectedcopiednodes{} }}\;. 
    \end{align}
    a relation which corresponds to a power law with exponent $\zeta = 1 + \frac{\brackets{k}}{\expectedcopiednodes{}}$.
    Expanding out $\expectedcopiednodes{}$ yields \Cref{eq:power-law-exponent} in the main text.
    Explicitly calculating this exponent requires the moments $\brackets{k}$, $\moment_{00}$, $\moment_{11}$, and $\moment_{01}$, which can be computed from the stationary joint distribution $\at{\jointdist}{\infty} \paren{k_0,k_1}$.

    \begin{figure}[t]
        \includegraphics[width=0.5\textwidth]{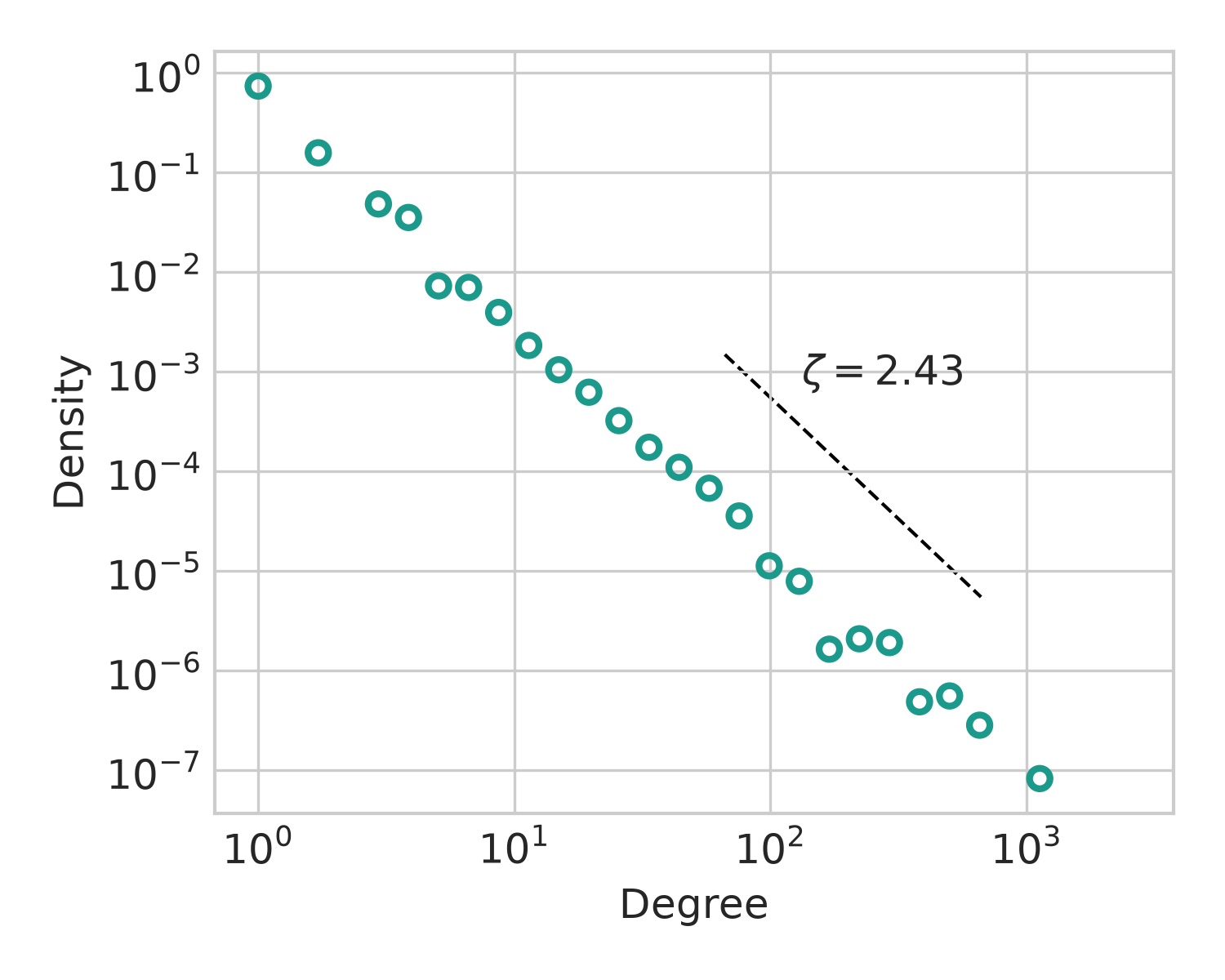}
        \caption{
            Degree distribution from model simulation of $10^4$ edges (points) and theoretical power law (line) with exponent given by \Cref{eq:power-law-exponent}. 
            The parameters in this experiment are $\same{\copyrate{}} = 0.7, \opp{\copyrate{}} = 0.3, \same{\extantrate{}} = 0.5, \opp{\extantrate{}} = \opp{\novelrate{}}3, \same{\novelrate{}} = 0.2, \opp{\novelrate{}} = 0.1$. 
            }
        \label{fig:degrees}
    \end{figure}

    \section{Additional Details on Stochastic Expectation-Maximization (SEM)} \label{sec:sufficient-statistics}
 
    \subsection{Sufficient Statistics}
        We describe the function $\vpsi$ from \Cref{sec:sem} which computes the conditional sufficient statistics for the parameters $\paramvector = (\same{\copyrate{}}, \opp{\copyrate{}}, \same{\extantrate{}}, \opp{\extantrate{}}, \same{\novelrate{}}, \opp{\novelrate{}})$ given an observed edge $e$, a candidate parent edge $f$, a candidate seed node $u$, and the node labels $\vz$. 
        Let $z = z_u$ be the label of the seed node $u$. 
        Let $e_z$ be the set of nodes in $e$ with label $z$, and let $f_z$ be the set of nodes in $f$ with label $z$. 
        Then, the sufficient statistics for the parameters $\paramvector$ \cite{degrootProbabilityStatistics2012}  are given by
        \begin{align*}
            \vpsi(e, f, u, \vz) = \begin{pmatrix}
                \abs{e_z \cap f_z} - 1 \\ 
                \abs{f_z} - 1 \\ 
                \abs{e_{\bar{z}} \cap f_{\bar{z}}} \\
                \abs{f_{\bar{z}}}  \\ 
                \abs{(e_z \setminus f_z) \cap \at{\nodes}{t}} \\ 
                \abs{(e_{\bar{z}} \setminus f_{\bar{z}})\cap \at{\nodes}{t}} \\
                \abs{e_z \setminus \at{\nodes}{t}} \\
                \abs{e_{\bar{z}} \setminus \at{\nodes}{t}}
            \end{pmatrix}^\top \;.
        \end{align*}
        As described in \Cref{sec:sem}, the vector $\sufficientstatsvec$ of expected sufficient statistics is computed as a weighted average of $\vpsi(e, f, u, \vz)$ over the belief distribution $\lik(f, u | e; \at{\hat{\paramvector}}{\ell})$ over the latent variables $f$ and $u$.
        Then, the function $g$ which computes the maximum likelihood estimates of the parameters $\paramvector$ from $\sufficientstatsvec$ is given by
        \begin{align*}
            g(\sufficientstatsvec) = \paren{
                \frac{s_1}{s_2}, 
                \frac{s_3}{s_4}, 
                s_5, 
                s_6, 
                s_7, 
                s_8
            }
             \;.
        \end{align*}
    
    \subsection{Learning Rate}

        We use a polynomially decaying learning rate for SEM with form $\atalg{\lr}{\epoch} = \atalg{\lr}{0} \cdot (1 - \frac{\epoch}{L})^{c}$ where $L$ is the iteration limit.
        The initial learning rate is set to $\atalg{\lr}{0} = 0.01$, and the decay rate $c = 0.9$.
        We initialized the vector of sufficient statistics $\atalg{\sufficientstatsvec}{0} = (1, 2, 1, 2, 0.5, 0.5, 0.5, 0.5)$.
        Other reasonable choices of hyperparameters did not substantially change the results of our experiments.

    \section{Synthetic SEM Parameter Values} \label{sec:SEM-appendix}
        In this section we give more details on the parameter values to demonstrate the performance of SEM on synthetic data in \Cref{fig:synthetic_sem}. 

    \begin{figure*}[t]
        \includegraphics[width=\textwidth]{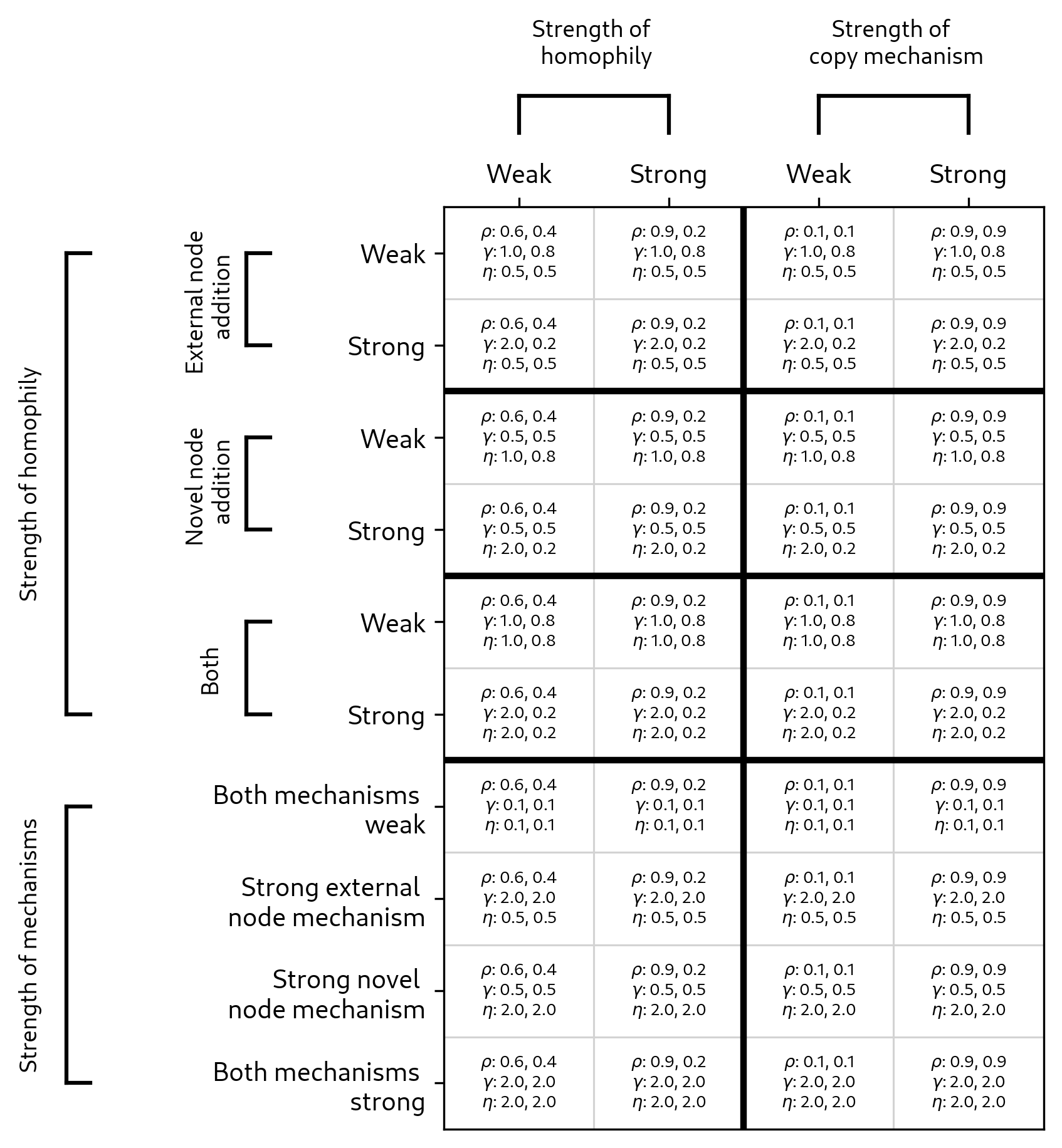}
        \caption{The parameter values used to generate each pixel of \Cref{fig:synthetic_sem}, where the first value in each pair is the $+$ (in-group) parameter and the second value is the $-$ (out-group) parameter.}
        \label{fig:synthetic-params}
    \end{figure*}

    \section{Approximation Argument for Simulated Anealing}\label{sec:simulated-annealing-approximation}

        This section provides further detail on the approximation to the likelihood described in \Cref{sec:community-detection} and used in our simulated annealing implementation. 
        The likelihood of an edge $e$ being generated in timestep $t$ can be written 
        \begin{widetext}    
            \begin{align}
                \lik(\newedge, \vz_\newedge'|\at{\vz}{t-1};\paramvector) = \sum_{\substack{\seededge \prec \newedge \\ u\in \seededge}} \lik(\newedge, \vz_\newedge' | \seededge, u, \at{\vz}{t-1}; \paramvector) \lik(\seededge,\focalnode | \at{\edges}{t-1}, \at{\vz}{t-1}; \paramvector) = \frac{1}{\at{m}{t}}\sum_{\seededge \prec \newedge}  \frac{1}{|\seededge|} \sum_{u \in \seededge \cap \newedge} \lik(\newedge, \vz_\newedge' | \seededge, u, \at{\vz}{t-1}; \paramvector) \;,
            \end{align} 
        \end{widetext}
        where $\at{m}{t}$ is the number of edges at time $t$.
        It is helpful to reindex this sum by the size of the intersection $|\seededge \cap \newedge|$ between $\seededge$ and $\newedge$. 
        \begin{align}
            \lik(\newedge, \vz_\newedge'|\vz;\paramvector) &= \sum_{h=0}^{|\newedge|} \sum_{\substack{\seededge \prec \newedge \\ |\seededge \cap \newedge| = k}} \frac{1}{\at{m}{t}}\sum_{u \in \seededge \cap \newedge} \frac{1}{|\seededge|} \lik(\newedge,\vz_\newedge' | \seededge, u, \vz; \paramvector) \;.
        \end{align}
        We will now argue that, when $n$ is large, the terms in this sum corresponding to the largest intersections (for which $h = \argmax _{\seededge \prec \newedge} |\seededge \cap \newedge|$) will dominate the likelihood, allowing us to neglect edges $\seededge$ whose intersection with $\newedge$ is smaller.

        Fix edge $\newedge$ generated at time $t$ and candidate parent edge $\seededge$. 
        We subscript $\newedge_z$, $\seededge_z$, and ${\at{\nodes}{t}}_z$ to denote the subsets of $\newedge$, $\seededge$, and $\at{\nodes}{t}$ with label $z$.
        We let ${\at{n}{t}}_z = |{\at{\nodes}{t}}_z|$ be the number of nodes of label $z$ at time $t$ and note that ${\at{n}{t}}_0 + {\at{n}{t}}_1 = \at{n}{t}$.
        Let $h_z = |(\newedge_z \setminus \seededge_z )\cap \nodes_z^t|$ be the number of extant nodes of label $z$ in $\newedge$.
        Then, the extant node addition process appears in $\lik(\newedge,\vz_\newedge' | \seededge, u, \vz; \paramvector)$ as a factor 
        \begin{align}
            E(h_0, h_1) = \frac{\poissonpdf{h_{z_u}}{\extantrate_+}\times\poissonpdf{h_{\bar{z}_u}}{\extantrate_-}}{\dbinom{{\at{n}{t}}_0 - |\seededge_{0}|}{h_{0}}\dbinom{{\at{n}{t}}_1 - |\seededge_{1}|}{h_{1}}} \;,
        \end{align}
        where the numerator describes the likelihood of adding fixed numbers of extant nodes of each label and the denominator describes the likelihood of selecting the \emph{specific set} of extant nodes from the set of all candidates. 
        Under the assumption that $\at{n}{t} \gg |\seededge|, |\newedge|$, we can estimate 
        \begin{align}
            \dbinom{{\at{n}{t}}_z - |\seededge_{z}|}{h_{z}} \approx \frac{({\at{n}{t}}_z)^{h_z}}{h_z!} \approx \frac{(c_z \at{n}{t})^{h_z}}{h_z!} \;,
        \end{align}
        where we have defined $c_z = \frac{{\at{n}{t}}_z}{\at{n}{t}}$ as the fraction of nodes with label $z$. 
        Then, absorbing terms that do not depend on $\at{n}{t}$ into a constant term $C(h_0, h_1)$, we have 
        \begin{align}
            E(h_0, h_1) &=  C(h_0, h_1) \at{n}{t}^{-h_0 - h_1} + O(\at{n}{t}^{-h_0 - h_1 - 1}) \\ 
                        &= C(h_0, h_1) \at{n}{t}^{-h} + O(\at{n}{t}^{-h - 1}) \;,
        \end{align}
        where $h = h_0 + h_1 = |(\newedge \setminus \seededge) \cap \at{\nodes}{t}|$ is the total number of extant nodes added to $\newedge$ and 
        \begin{align}
            C(h_0, h_1) &= \frac{\poissonpdf{h_{z_u}}{\extantrate_+}\times\poissonpdf{h_{\bar{z}_u}}{\extantrate_-}}{\frac{(c_0)^{h_0}}{h_0!}\frac{(c_1)^{h_1}}{h_1!}} \;.
        \end{align}
        It follows that $\lik(\newedge,\vz_\newedge' | \seededge, u, \vz; \paramvector)$ is of order $O(\at{n}{t}^{-h})$ as $\at{n}{t}$ grows large.
        Terms corresponding to choices of $\seededge$ for which $h$ is larger (of which there may be up to approximately $\at{m}{t}$) decay quickly as $\at{n}{t}$ grows large. 
        We therefore have 
        \begin{widetext}
            \begin{align}
                \lik(\newedge,\vz_\newedge'|\vz;\paramvector) &= \sum_{\substack{\seededge \prec \newedge \\ |\seededge \cap \newedge| = h'}} \frac{1}{\at{m}{t}}\sum_{u \in \seededge \cap \newedge} \frac{1}{|\seededge|}  \lik(\newedge,\vz_\newedge' | \seededge, u, \vz; \paramvector)  + \at{m}{t}O(\at{n}{t}^{-h'-1}) \label{eq:sa-approx}
            \end{align}
        \end{widetext}
        where $h' = \min_{\seededge \prec \newedge} |(\newedge \setminus \seededge) \cap \at{\nodes}{t}|$ is the minimum number of extant nodes added to $\newedge$ across all choices of $\seededge$.
        
        This argument suggests that, when $\at{n}{t}$ is relatively large in relation to $\at{m}{t}$, the likelihood of $\newedge$ is dominated by terms corresponding to choices of $\seededge$ for which the number of extant nodes added to $\newedge$ is small.
        There are two limitations to this argument in the context of empirical data sets. 
        First, it may be the case that only a small number of candidate parent edges $\seededge$ have large intersections with $\newedge$ , which may lead to a poor approximation of the likelihood when we evaluate only a small number of corresponding terms.
        Second, several of our data sets have $\at{m}{t}$ large in relation to $\at{n}{t}$, making it dubious to neglect the error term. 
        We therefore introduce two variations to our approximation heuristic: 
        \begin{itemize}
            \item \textbf{Top $j$ heuristic}: Rather than evaluate only terms for $f$ for which the number of extant nodes $h$ to be added is minimal, we evaluate terms for the top $j$ choices of $\seededge$ with the smallest values of $h$.
            \item \textbf{Batch normalization}: Rather than normalizing by $\frac{1}{\at{m}{t}}$ in \cref{eq:sa-approx}, we normalize by the number of  choices for $\seededge$ that we evaluate for a specific $\newedge$. 
            This may be $j$ under the top $j$ heuristic, or it may be the number of $\seededge$ choices with $h$ equal to the minimum $h'$ if we do not use the top $j$ heuristic. 
            Letting $\mathcal{F}_\newedge$ be the set of seed edges $\seededge$ evaluated for a given $\newedge$ under the top $j$ or similar heuristic, our approximation is 
            \begin{align}
                \tilde{\lik}(\newedge,\vz_\newedge'|\vz;\paramvector) &=\frac{1}{|\mathcal{F}_\newedge|}\sum_{\substack{\seededge \in \mathcal{F}_\newedge}} \sum_{u \in \seededge \cap \newedge} \frac{1}{|\seededge|}  \lik(\newedge,\vz_\newedge' | \seededge, u, \vz; \paramvector)\;. 
            \end{align}
            The approximate likelihood of the complete data set is then 
            \begin{align}
                 \tilde{\lik} (\hypergraph, \vz; \paramvector )&= \prod_{\newedge \in \edges}\tilde{\lik}(\newedge,\vz_\newedge'|\vz;\paramvector)\;. \label{eq:sa-approx-2}
            \end{align}
        \end{itemize}
        It is important to note that the batch normalization \cref{eq:sa-approx-2} tends to overestimate the likelihood, since we have restricted ourselves to an average of high-probability seed-edges. 
        For the purposes of simulated annealing, this is not a major problem since we are only interested in the relative likelihood of different labelings $\vz$ and not the absolute scale of likelihood.

\raggedbottom

    \section{Computational Notes on Simulated Annealing}\label{sec:simulated-annealing-algorithm-spec}
        We give several additional details on our simulated annealing method (\Cref{algocf:simulated_annealing}) described in \Cref{sec:community-detection}. 
        Our custom learning schedule uses the first epoch to set the absolute scale of the acceptance probability decay, and we therefore describe this epoch as a separate algorithm (\Cref{algocf:simulated_annealing_first_epoch}).
        We then proceed to the main simulated annealing algorithm (\Cref{algocf:simulated_annealing}), which uses the scale of acceptance probability decay $\sigma_\hypergraph$ computed in the first epoch to set the acceptance probability for label flips in subsequent epochs.
        We only allow the probability to begin to decay after the first four epochs in order to encourage exploration of the label space. 
        
        \begin{algorithm}[H]
            \begin{algorithmic}
                \caption{\texttt{SimulatedAnnealingFirstEpoch}$(\hypergraph, \paramvector)$}\label{algocf:simulated_annealing_first_epoch}
                \Require{Hypergraph $\hypergraph$ with $n$ nodes, parameters $\paramvector$}
                \Ensure{Estimate $\sigma_\hypergraph$ for scale of acceptance probability decay, updated label set $\hat{\vz}$}
                \State Sample $\hat{\vz} \sim \text{Uniform}(\{0,1\})^{n}$ \Comment{initialization}
                \State $\vdelta \gets \vzero \in \R^n$ \Comment{initialize vector of log-likelihood differences}
                \For{$\step = 1$ to $n$}
                    \State Sample $j$ $\sim$ Uniform($\{1,...,n\}$)
                    \State $\hat{\vz}^\prime$ $\gets$ $\hat{\vz}$
                    \State $z^\prime_j\gets1-z^\prime_j$
                    \State $\vdelta_\tau \gets \log \tilde{\lik}(\hypergraph, \hat{\vz}^\prime ; \paramvector) - \log \tilde{\lik}(\hypergraph, \hat{\vz} ; \paramvector)$ \Comment{populate $\vdelta$ with log-likelihood differences}
                \EndFor
                \State $\sigma_\hypergraph\gets \sigma(\vdelta)$
                \State \Return $\sigma_\hypergraph$, $\hat{\vz}$
            \end{algorithmic}
        \end{algorithm}

        Here, $\sigma(\vdelta)$ is the square root of the empirical variance
        \begin{align}
            \sigma(\vdelta) = \sqrt{\frac{\sum_{i = 1}^n \delta_i^2 - (\frac{1}{n}\sum_{i=1}^n \delta_i)^2}{n-1}}\;.
        \end{align}

        It is possible to store computational components required to evaluate $\tilde{\cL}(\vz)$ in a data structure that allows us to compute the change in likelihood from the prior state without recomputing the entire likelihood.
        Although this does not result in an asymptotic improvement, experimentally we observe a roughly 30-fold reduction in compute time on our empirical data sets at the expense of increasing memory requirements.

    \section{Simulated Annealing Illustration}

        \begin{figure*}
            \includegraphics[width=1.0\textwidth]{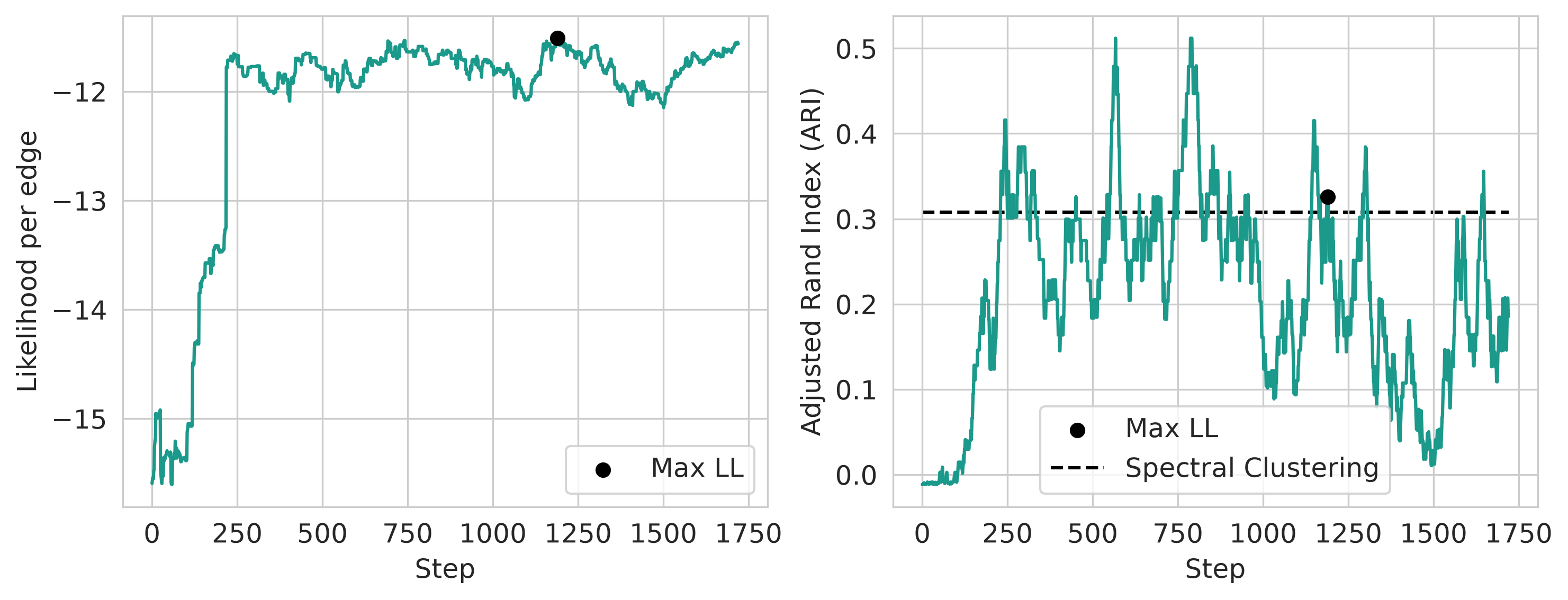}
            \caption{
                Illustrative run of simulated annealing for community detection in a hypergraph generated from our model.
                The dashed line on the right shows the ARI achieved by Laplacian spectral clustering applied to the weighted dyadic graph projection.  
                Solid point gives the highest likelihood and corresponding ARI achieved by simulated annealing over the course of the run. 
            }
        \end{figure*}
        We provide simple illustration of a single run of simulated annealing as described in \Cref{algocf:simulated_annealing_first_epoch,algocf:simulated_annealing} for community detection in a hypergraph generated from our model.
        In this simulation, we generated a simple hypergraph according to our model with \simannviznumnodes{} nodes and \simannviznumedges{} edges.
        Running simulated annealing for \simannviznumsteps{} steps, we obtained a maximum log-likelihood of \simannvizmaxll{} nats per edge with an Adjusted Rand Index (ARI) against ground-truth of \simannvizmaxllari{}. 
        Laplacian spectral clustering as implemented in the Python package \texttt{networkx} \cite{networkx} obtained an ARI of \simannvizspectralari{}.
        The parameters for the model used to generate the hypergraph are 
        \begin{align*}
            \paramvector = (\same{\copyrate{}}, \opp{\copyrate{}}, \same{\extantrate{}}, \opp{\extantrate{}}, \same{\novelrate{}}, \opp{\novelrate{}}) = (\simannvizsamecopyrate{}, \simannvizoppcopyrate{}, \simannvizsameextantrate{}, \simannvizoppextantrate{}, \simannvizsamenovelrate{}, \simannvizoppnovelrate{})\;.
        \end{align*}
        For the purposes of this illustration, we perform community detection under the Nishimori condition, using a guess $\hat{\paramvector}$ equal to the true parameters $\paramvector$ used to generate the hypergraph.

    \end{document}